\documentclass[preprints,article,accept,pdftex,moreauthors]{Definitions/mdpi} 

\firstpage{1} 
\pubvolume{1}
\issuenum{1}
\articlenumber{0}
\pubyear{2024}
\copyrightyear{2024}
\datereceived{ } 
\daterevised{ } 
\dateaccepted{ } 
\datepublished{ } 
\hreflink{https://doi.org/} 

\newcommand{\beq}{\begin{eqnarray}}
\newcommand{\eeq}{\end{eqnarray}}

\newcommand{\Slash}[1]{{\ooalign{\hfil/\hfil\crcr$#1$}}}
\newcommand{\tr}{{\rm tr}}

\newcommand{\Nc}{N_{\rm c}}
\newcommand{\Nf}{N_{\rm f}}

\newcommand{\lqcd}{\Lambda_{\rm QCD}}
\newcommand{\vp}{\vec{p}}
\newcommand{\vq}{\vec{q}}

\newcommand{\vl}{\vec{l}}

\newcommand{\la}{\langle}
\newcommand{\ra}{\rangle}

\newcommand{\para}{\parallel}

\newcommand{\calL}{\mathcal{L}}

\newcommand{\calB}{\mathcal{B}}
\newcommand{\calC}{\mathcal{C}}

\newcommand{\calM}{\mathcal{M}}

\newcommand{\calP}{\mathcal{P}}

\newcommand{\rmd}{\mathrm{d}}
\newcommand{\rmi}{\mathrm{i}}
\newcommand{\rme}{\mathrm{e}}

\Title{Isospin QCD as a laboratory for dense QCD }

\TitleCitation{Title}

\Author{Toru Kojo $^{1}$\orcidA{}, Daiki Suenaga $^{2,3}$ and Ryuji Chiba $^{1}$}

\AuthorNames{Firstname Lastname, Firstname Lastname and Firstname Lastname}

\AuthorCitation{Kojo, T.; Suenaga, D.; Chiba, R.}

\address{$^{1}$ \quad Department of Physics, Tohoku University, Sendai 980-8578, Japan\\
$^{2}$ \quad Kobayashi-Maskawa Institute for the Origin of Particles and the Universe, Nagoya University, Nagoya, 464-8602, Japan\\
$^{3}$ \quad Research Center for Nuclear Physics, Osaka University, Ibaraki 567-0048, Japan}

\corres{Correspondence: toru.kojo.b1@tohoku.ac.jp}

\abstract{
QCD with the isospin chemical potential, $\mu_I$, is a useful laboratory to delineate the microphysics in dense QCD.
To study the quark-hadron-continuity 
we use a quark-meson model that interpolates hadronic and quark matter physics at microscopic level.
The equation of state is dominated by mesons at low density but taken over by quarks at high density.
We extend our previous studies with two-flavors to the three-flavors case to study the impact of the strangeness 
which may be brought by kaons $(K_+, K_0) = (u\bar{s}, s\bar{d})$ and the U$_A$(1) anomaly.
In the normal phase the excitation energies of kaons are reduced by $\mu_I$ 
in the same way as hyperons in nuclear matter at finite baryon chemical potential.
Once pions condense, kaon excitation energies increases as $\mu_I$ does.
Moreover, strange quarks become more massive through the U$_A$(1) coupling to the condensed pions.
Hence at zero and low temperature the strange hadrons and quarks are highly suppressed.
The previous findings in two-flavor models, sound speed peak, negative trace anomaly, gaps insensitve to $\mu_I$, 
persist in our three-flavor model and remain consistent with the lattice results to $\mu_I \sim 1$ GeV.
We discuss the non-perturbative power corrections and quark saturation effects
as important ingredients to understand the crossover equations of state measured on the lattice.
}


\begin{document}

\section{Introduction}

Recently there have been increasing attentions to 
two-color QCD (QC$_2$D) \cite{Kogut:1999iv,Kogut:2000ek,Iida:2024irv,Iida:2019rah,Iida:2020emi,Boz:2019enj,Boz:2013rca,Cotter:2012mb,Hands:2011ye,Astrakhantsev:2020tdl,Bornyakov:2020kyz,Muroya:2002ry,Suenaga:2022uqn,Suenaga:2023xwa,Kawaguchi:2024iaw,Sun:2007fc,Brauner:2009gu,Strodthoff:2013cua,Strodthoff:2011tz} 
or QCD at finite isospin but zero baryon densities (isospin QCD, QCD$_I$ in short) \cite{Abbott:2024vhj,Abbott:2023coj,Brandt:2022hwy,Son:2000xc,Son:2000by,Splittorff:2000mm,GomezNicola:2022asf,Lu:2019diy}.
In these theories lattice simulations are doable without the sign problem.
Confronting theories with lattice results should provide us with useful insights into QCD matter at high baryon density ($n_B$), see, e.g., reviews \cite{Fukushima:2010bq,Oertel:2016bki,Baym:2017whm,Kojo:2020krb,Vuorinen:2024qws}.

Equations of state (EOS) of dense matter have one-to-one correspondence with the mass-radius ($M$-$R$) relations of neutron stars.
One of important indications from neutron star observations and nuclear constraints is that QCD matter is soft around nuclear saturation density $n_0 \simeq 0.16\, {\rm fm}^{-3}$, 
but rapidly becomes stiff around $n_B = 2$-$4n_0$ \cite{Drischler:2020fvz,Han:2022rug}.
This density for this stiffening is smaller than the density where baryons of the radius $0.5$-$0.8$ fm spatially overlap.
Meanwhile nuclear many-body calculations become problematic for $n_B \gtrsim 1.5$-$2n_0$.
Observationally rapid stiffening is supported by small variation (or even increase) in radii from $1.4M_\odot$ to $2.1M_\odot$ neutron stars.
If EOSs stiffen only gently as in typical hadronic EOS with many-body forces, 
the radii of $\simeq 2M_\odot$ neutron stars would be substantially smaller than $1.4M_\odot$ neutron stars by $\sim 1$ km \cite{Kojo:2021wax}. 
Such contrast in $1.4M_\odot$ to $2.1M_\odot$ radii may be also studied by gravitational waves from binary neutron star mergers,
see, e.g., Refs.~\cite{Huang:2022mqp,Fujimoto:2022xhv,Kedia:2022nns,Bauswein:2018bma}.

To study the interplay between hadronic and quark matter, it is crucial to understand the rapid stiffening at microscopic level.
Several theoretical studies \cite{McLerran:2018hbz,Jeong:2019lhv,Kojo:2021ugu,Fujimoto:2023mzy} suggest that such rapid stiffening around $2$-$4n_0$ is triggered by quark degrees of freedom,
whether or not quarks are confined or deconfined.
Assuming a model in which quarks remain confined in baryons, 
quarks can still occupy states with a certain probability and eventually affect baryons through the quark Pauli blocking \cite{Kojo:2021ugu,Fujimoto:2023mzy}.
Such quark constraint becomes substantial even before baryons overlap.
After quark states at low momentum are saturated (``quark saturation'' \cite{Kojo:2021ugu}), the quark Fermi sea begins to form with the diffused Fermi surface whose thickness is $\simeq 200$-$300$ MeV. 
The quark saturation forces matter of non-relativistic baryons to change into that of relativistic baryons or quarks, 
driving the rapid growth in pressure but modest change in the energy density \cite{Kojo:2021ugu}.
The quark momentum distribution at finite density has been manifestly computed for QCD in two-dimensional QCD \cite{Hayata:2023pkw}.

The quark saturation and the associated stiffening may occur also in QC$_2$D and QCD$_I$, 
although baryons are replaced with diquark baryons and mesons, respectively.
In these theories the Bose-Einstein Condensation of diquarks or mesons occur at the onset of matter;
(composite) bosons occupy the zero momentum state.
At finite chemical potential exceeding the mass threshold,
the amplitude of condensates would grow indefinitely unless some sort of repulsive forces temper the growth of the amplitude.
For theories of elementary bosons we do not have definite rationals why such repulsion should exist,
while, for theories of bosons made of fermions, indefinite growth of boson amplitudes would violate the Pauli exclusion principle.
Hence, irrespective to the details of interactions, effective repulsions among bosons must emerge.

For isospin QCD there have been many studies, see Ref.~\cite{Mannarelli:2019hgn} for recent good summary to 2019.
Recently isospin QCD attracts renewed attention as a laboratory to study concepts proposed for neutron star 
physics \cite{Chiba:2023ftg,Chiba:2024cny,Fujimoto:2023mvc} 
or to derive some constraints on neutron star EOS \cite{Cohen:2003ut,Navarrete:2024zgz,Fujimoto:2023unl,Moore:2023glb}.
In Refs.~\cite{Chiba:2023ftg,Chiba:2024cny},
we used a two-flavor quark-meson model to discuss the rapid stiffening and related microphysics.
This model is renormalizable and includes mesons and quarks \cite{Adhikari:2016eef,Ayala:2023cnt}.
At low density the EOS is dominated by mesons while, at high density, quarks dominate. 
In quark matter region mesonic degrees of freedom as condensates remain near the quark Fermi surface.

In this paper, after adding some supplements to our previous two-flavor studies,
we then extend the analyses to three-flavors.
Although a strange quark does not have isospin,
hadronic strange particles, such as kaons, $\eta, \eta'$, and so on,
contain $u$,$d$-quarks and can be affected by the isospin density. 
Indeed the excitation energies of $(K_+, K_0) = (u\bar{s}, s\bar{d})$
decrease with increasing $\mu_I$ and would eventually condense unless other particles in the system repel such kaons.
Another effect of interest is the impact of U$_A$(1) anomaly that affects the strangeness in the Dirac sea \cite{Gao:2022klm,Minamikawa:2023eky}.
At finite density, the reduction of the effective $u$,$d$ quark masses softens the chiral symmetry breaking in the strange quark sector,
while the appearance of the pion condensates can enhance the strange quark mass by few hundred MeV.
This continues until the medium screening cutoff such effects.
Similar coupling, diquark-to-chiral order parameters, has been discussed in the context of 
quark-hadron-crossover \cite{Hatsuda:2006ps,Yamamoto:2007ah,Hatsuda:2008is,Zhang:2008ima}.
Thus the strange particles are useful probes to diagnose various medium effects in dense matter.

In this paper we use a quark-meson model to discuss a quark-hadron crossover \cite{Schafer:1998ef,Masuda:2012kf,Masuda:2012ed,Masuda:2015kha,Tajima:2022iqw,Ma:2019ery}. 
In Sec.~\ref{sec:QM_model} we first discuss a picture behind this model to inform readers of what physics we try to describe.
We will use the $1/\Nc$ expansion to classify the effects caused by different degrees of freedom \cite{tHooft:1973alw,Witten:1979kh}.
In Sec.~\ref{sec:Lagrangian} we present a quark-meson model including the strangeness and the U$_A$(1) anomaly effects.
In Sec.~\ref{sec:tree} we study the effective potential and meson spectra at tree level.
In Sec.~\ref{sec:quark_loop} we include quarks to change the structure of the theory and discuss how it leads to the trend consistent with the lattice results.
In Sec.~\ref{sec:meson_poles} we mention our parameter fixing through meson poles in vacuum with quark-loop corrections.
In Sec.~\ref{sec:eos} we examine EOS at zero temperature and its relationship with the microphysics.
Sec.~\ref{sec:summary} is devoted to summary.

\section{A quark-meson model at finite density: outline}
\label{sec:QM_model}


\begin{figure}[t]
\vspace{-1.0cm}
\begin{center}
\includegraphics[width=10. cm]{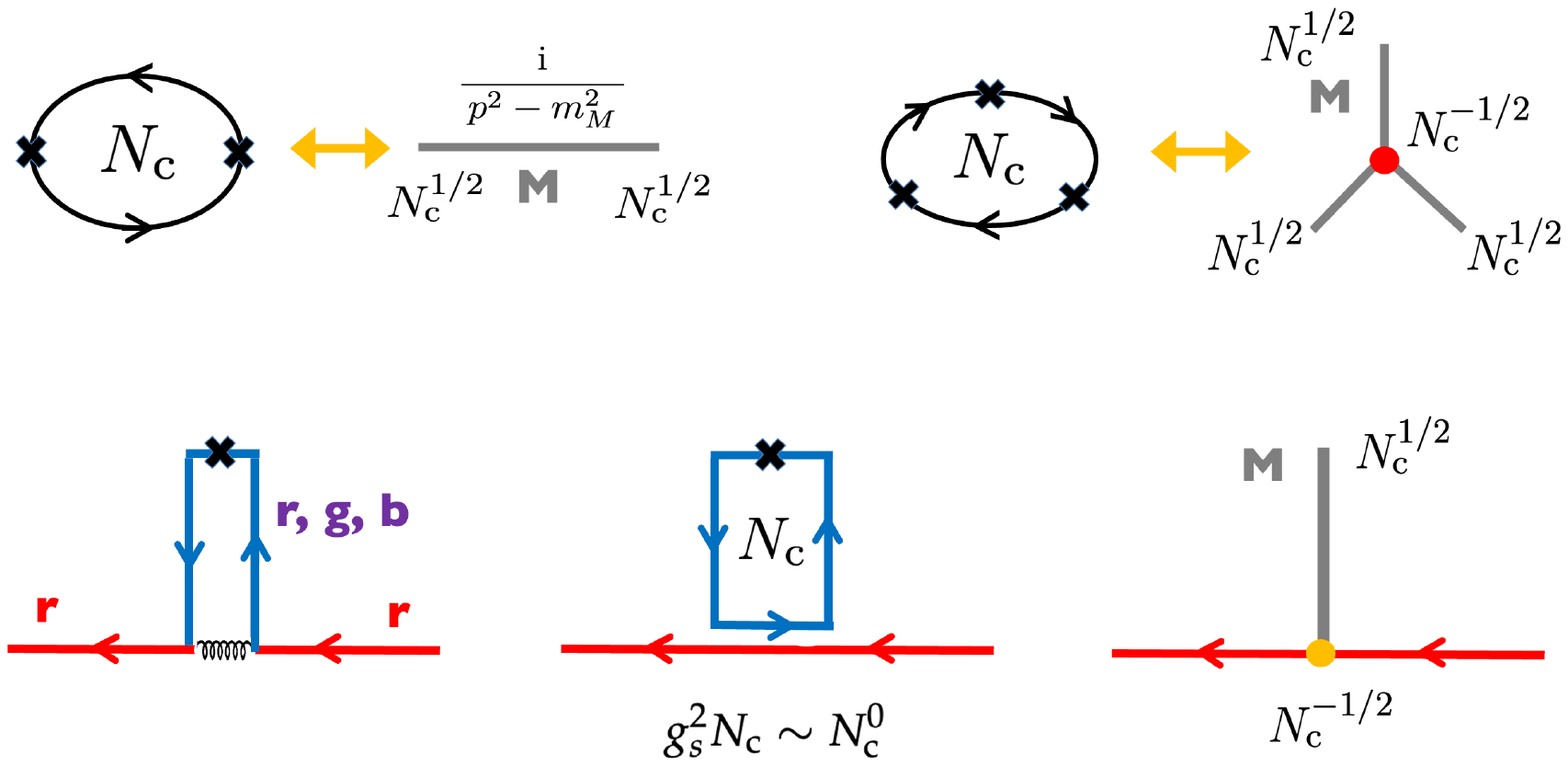}
\end{center}
\vspace{-1.5cm}
\caption{(upper) Some examples of the correspondence between quark-gluon, color line graphs, and hadronic graphs. 
The coupling between quark bilinear operators and mesonic state must be $\sim \Nc^{1/2}$ for the consistency with the color line graphs. 
Meson three point vertices must be $\sim \Nc^{-1/2}$.
(lower) The quark-meson vertices in quark-gluon, color-line, and quark-meson representations.
The quark-meson coupling must be $\sim \Nc^{-1/2}$ to be consistent with the color line representation.
}
\label{fig:graph_q-m_coupling}
\end{figure}   

We begin with the $\Nc$ counting of a quark-meson coupling.
A diagrammatic representation is given in Fig.~\ref{fig:graph_q-m_coupling}.
To derive the $\Nc$ counting of hadronic parameters we match the $\Nc$ counting in a color line graph and a hadronic graph. 
For instance, from a mesonic correlator with the magnitude of $\Nc$, 
one can conclude that a quark bilinear operator couples to a mesonic state with the strength $\Nc^{1/2}$.
Using this estimate and repeating the similar matching,
one can deduce that the meson-meson interaction is $\sim \Nc^{-1/2}$ for three meson vertices and $\sim \Nc^{-1}$ for four meson vertices, and so on.
Similarly, for a quark meson coupling graph with the amplitude of $\sim g_s^2 \Nc \sim \Nc^0$, 
one can conclude that the quark-meson coupling is $\Nc^{-1/2}$.
While we consider only a single gluon exchange, including more gluons introduces $(g_s^2 \Nc)^n \sim 1$ so that the counting is not affected.

\begin{figure}[t]
\vspace{-1.0cm}
\begin{center}
\includegraphics[width=10. cm]{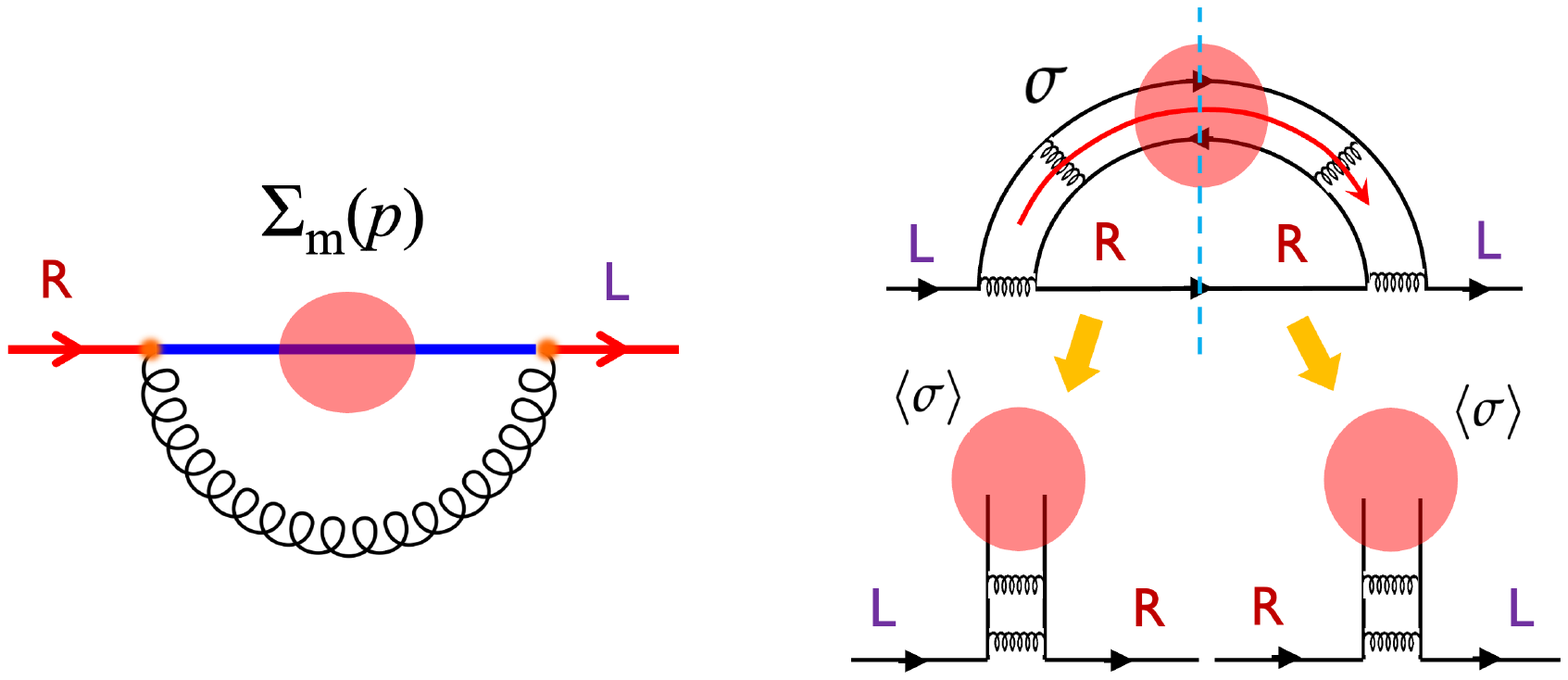}
\end{center}
\vspace{-1.5cm}
\caption{ The mass self-energy graphs with (left) gluon and (right) meson loops.
The gluon loop graph is $\sim g_s^2 \Nc \sim \Nc^0$.
The meson loop graph is $ \sim g^2 \sim g_s^4 \Nc \sim \Nc^{-1}$ except when the meson condenses.
With meson condensates of $\la \sigma \ra \sim f_\pi \sim \Nc^{1/2}$, the graphs with condensate can represent (some part of) the gluon loop graphs in the leftmost panel.
}
\label{fig:mass_self-energy}
\end{figure}   

Below we assume that this quark-meson coupling, $g$, is dominated by soft gluons for which the quark-gluon coupling $g_s$ is large
and characterizes the size of the quark-meson coupling.
The soft gluon exchanges occur indefinitely within the produced mesons.
This sort of process cannot be represented by perturbative treatments at finite order.
The purpose of manifestly including mesonic degrees of freedom is to replace or parametrize these process in a book keeping manner.

The quark self-energy with gluon loops (leftmost panel, Fig.~\ref{fig:mass_self-energy}) are $g_s^2 \Nc \sim \Nc^0$ yielding the mass of $\sim \lqcd$.
The factor $\Nc$ amplification occurs as there is a color loop in color line representations.
In contrast, the self-energy with mesonic loops are $g^2 \sim g_s^4 \Nc \sim \Nc^{-1}$ and suppressed (rightmost panel, Fig.~\ref{fig:mass_self-energy}).
There is an exception, however.
When mesons form condensates, the self-energies originating from meson loops can be amplified to $\Nc^0$.
The condensate contains indefinite numbers of $q\bar{q}$ pairs.
In the case of chiral condensates, condensed $\bar{q}_L q_R$ or $\bar{q}_R q_L$ pairs transform $q_L (q_R)$ into $q_R (q_L)$ generating the mass self-energies.
If we neglect changes in the number of $\bar{q} q$ pairs in the condensate,
a quark propagator with meson loops can be factorized into a product of quark propagators with a background meson condensate.
Since quark propagators with the meson background already contain graphs with many soft gluons, 
to avoid double counting
we do not manifestly include the quark self-energy graphs with soft gluon loops.
Meanwhile graphs with hard gluons may be added separately for more complete exploration of the phase space
in the intermediate states.

The reduction of meson masses can be important even before mesons condense.
Such precursory effects are important near the phase boundaries and have been studied 
in the context of color-superconductivity \cite{Nishimura:2022mku,Kitazawa:2005vr,Kitazawa:2001ft}.

We include ``mesons'' as a representative of $q \bar{q}$ propagating graph with indefinite numbers of soft gluon exchanges.
Using meson degrees of freedom at high density might look unnatural,
but models manifestly including mesons can be dynamical reduced to pure quark descriptions;
the dissociation and structural changes of these mesons into quarks can be described by inserting quark loops 
in the meson propagators.
Thus unwanted contributions can be canceled by including quark loops \cite{Blaschke:2013zaa}.

At high density, quarks typically have large momenta but the excitation energies can be small, $\sim E-\mu \ll \mu $ ($\mu$: chemical potential) near the Fermi surface.
These soft quarks and soft gluons may keep quark-meson couplings as strong as in vacuum unless screening processes cut off soft gluons.
For S-wave condensates, a quark with $\vp$ and an antiquark with $-\vp$ make a pair in which soft gluons are exchanged.
For a pion condensed phase in isospin QCD, 
these soft gluons should be largely unscreened because the condensate is color-singlet and colored excitations are gapped.
These unscreened gluon propagators are supported by lattice simulations for two-color QCD \cite{Boz:2018crd,Bornyakov:2020kyz,Kojo:2014vja,Suenaga:2019jjv,Kojo:2021knn,Contant:2019lwf}.

\begin{figure}[t]
\vspace{-1.0cm}
\begin{center}
\includegraphics[width=10. cm]{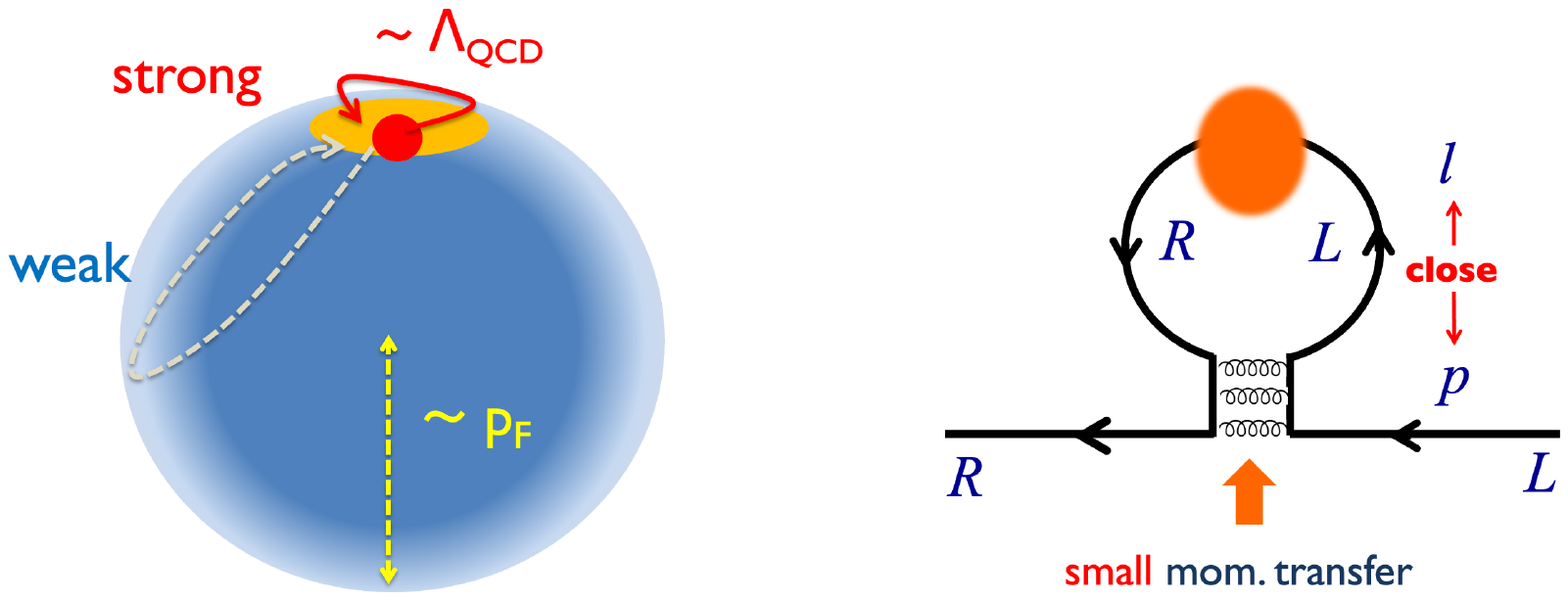}
\end{center}
\vspace{-1.5cm}
\caption{ The gap created by soft gluons. 
The self-energies caused by soft gluons are not very sensitive to the size of the Fermi surface because quarks fluctuate only within small domain.
The produced gap is characterized by soft gluons and is supposed to be $\sim \lqcd$, until hard gluon processes dominate over soft gluon processes.
}
\label{fig:mass_gap}
\end{figure}   

With soft gluons being unscreened, the size of the gap is $\sim \lqcd$ or can be even larger.
To see this, we parametrize the soft gluon exchange forces as
\beq
D_g (q) = \Theta \big( \Lambda - |\vq| \big) \frac{\, c \,}{\, \Lambda^2 \,} \,,~~~~~~ c \sim O(1) \,,
\eeq
where $c/\Lambda^2$ is the typical strength for a soft momentum transfer $|\vq| < \Lambda$.
This model has been used for quarkyonic chiral spirals \cite{Kojo:2009ha,Kojo:2010fe,Kojo:2011cn} and QCD in a magnetic field \cite{Kojo:2012js,Kojo:2014gha,Hattori:2015aki}
to yield the gap equation local in momentum space (Fig.~\ref{fig:mass_gap}).
The resultant gap is $\sim \Lambda$, and we expect the same happens for the soft gluonic part in isospin QCD.

Now the gap equation reads
\beq
\Delta 
~\sim~  
\frac{\, c \,}{\, \Lambda^2 \,}\int_{\vl} \Theta \big( \Lambda^2 - |\vl -\vp|^2 \big) \frac{\, \Delta \,}{\, \sqrt{ \big( E (\vl) - \mu_I \big)^2 + \Delta^2 \,} \,} \,.
\eeq
Decomposing the loop momentum as $\vl = \big( |\vp| + \delta l_{\para} \big) {\bf e}_{p} + \vl_\perp$ with $|\vp| \simeq \mu_I$, 
the conservative estimate for the range is $|\delta l_\para| \lesssim \Lambda$ and $|\vl_\perp| \lesssim \Lambda$.
Then one can simplify the gap equation into ($v_F$: Fermi velocity)
\beq
1 
~\sim~  
\frac{\, c \,}{\, (2\pi)^3 \Lambda^2 \,} \int_{ -\Lambda}^{\Lambda} \rmd \delta l_\para \int_{ |\vl_\perp| \lesssim \Lambda}  \rmd \vl_\perp
\frac{\, 1 \,}{\, \sqrt{ \big( v_f \delta l_\para )^2 + \frac{\, \vl_\perp^2 \,}{\, 2\mu_I \,} + \Delta^2 \,} \,} \,.
\eeq
When $\Lambda \ll \mu_I $ holds, we can omit the $\vl_\perp^2/\mu_I$ term in the range of integral to factorize the $\vl_\perp$ integral. 
At this stage the equation loses the $\mu_I$-dependence and so does the solution.
Further calculation reads
\beq
1 
~\sim~  
\frac{\, c \,}{\, 2 \pi^2 \,} \int_{ 0}^{\Lambda} \rmd \delta l_\para 
\frac{\, 1 \,}{\, \sqrt{ \big( v_f \delta l_\para )^2  + \Delta^2 \,} \,}
~\sim~ 
\frac{\, c \,}{\, 2 \pi^2 \,} \ln \frac{\, \Lambda \,}{\, \Delta \,} \,.
\eeq
Thus the gap is $\Delta \sim \Lambda \rme^{ - 2\pi^2 /c }$.
For a stronger coupling the gap $\Delta$ grows to $\Lambda$.

The present arguments show that,
even at large $\mu$ or large quark Fermi momenta, non-perturbative phenomena characterized by $\lqcd$ are still possible.
This continues until screening effects set in.
When $\mu_I$ is sufficiently large, we should also add hard gluon contributions which can be treated within weak coupling method;
this hard contributions are sensitive to the phase space around the Fermi surface and hence to $\mu_I$.
As we see later the gap (which is insensitive to $\mu_I$) adds the {\it power corrections}, $\sim + \mu_I^2 \Delta^2 \sim \mu_I^2 \lqcd^2$, to the pressure $P(\mu_I)$.
This power corrections, not capable within perturbative computations, 
have played important roles in QCD phenomenology \cite{Novikov:1977dq,Shifman:1978bx,Shifman:1978by}
and also
seem essential to describe the lattice simulation data at large density \cite{Chiba:2023ftg,Chiba:2024cny}.

%

\section{A quark-meson model: mean field treatments}
\label{sec:Lagrangian}

Now we consider a practical description of a three-flavor quark-meson model. The Lagrangian is 
\beq
\calL = \calL_{\rm Y} + \calL_{\rm kin}^M - V_{\rm sym} - V_{\rm SB} - V_{\rm anom} \,.
\eeq
The quark with the Yukawa interaction takes the form
\begin{align}
\calL_{\rm Y}
&= \bar{q} \big[ \rmi \Slash{\partial} + \mu_I \tau_3 \gamma_0 - g \big( \sigma_a + \rmi \gamma_5 \pi_a  \big) \lambda_a  \big] q  
\notag \\
&= \bar{q} \big[ \rmi \Slash{\partial} + \mu_I \tau_3 \gamma_0  \big] q 
- g \phi_a \big[  \bar{q}_L  \lambda_a q_R \big] - g \phi^*_a \big[ \bar{q}_R  \lambda_a  q_L \big]
 \,,
\end{align}
where $\phi_a = \sigma_a + \rmi \pi_a$. We have scalar and pseudoscalar flavor nonets.
In our notation the isospin chemical potential is $\mu_I = \mu_u = - \mu_d$, coupled to $ \bar{q} \gamma_0 \tau_3 q$ 
rather than the isospin density operator $ \bar{q} \gamma_0 \frac{\, \tau_3 \,}{2} q$ in the conventional sense.

The mesonic Lagrangian with the isospin chemical potential is
\begin{align}
	\calL^M_{\rm kin}=&\, \sum_{a=0,3,8} \frac{1}{\, 2 \,} | \partial_\mu \phi_a |^2 \notag\\
		&+ (\partial_\mu + 2 \rmi \mu_I\delta_\mu^0) a_0^+ (\partial^\mu - 2 \rmi \mu_I\delta^\mu_0 ) a_0^-
			+ (\partial_\mu + 2 \rmi \mu_I\delta_\mu^0) \pi^+ (\partial^\mu - 2 \rmi \mu_I\delta^\mu_0 ) \pi^- \notag \\
		& + (\partial_\mu +  \rmi \mu_I\delta_\mu^0) \kappa^+ (\partial^\mu -  \rmi \mu_I\delta^\mu_0 ) \kappa^- 
			+ (\partial_\mu +  \rmi \mu_I\delta_\mu^0) K^+ (\partial^\mu -  \rmi \mu_I\delta^\mu_0 ) K^- \notag\\
		& + (\partial_\mu +  \rmi \mu_I\delta_\mu^0) \bar{\kappa}^0 (\partial^\mu -  \rmi \mu_I\delta^\mu_0 ) \kappa^0 
			+ (\partial_\mu +  \rmi \mu_I\delta_\mu^0) \bar{K}^0 (\partial^\mu -  \rmi \mu_I\delta^\mu_0 ) K^0 \,,
\end{align}
where iso-triplet $(\sigma_{a=1-3}, \pi_{a=1-3}) = (a_0, \pi)$ and two iso-doublets $(\sigma_{a=4-7}, \pi_{a=4-7}) = (\kappa, K)$ couple to the isospin chemical potential.
The field normalization for isospin charged fields are  $a_0^\pm = (\sigma_1 \pm \rmi \sigma_2)/\sqrt{2}$, and so on.
Since $\tau_3$ commutes with the isospin rotation operator around the $I_3$-axis, the U(1)$_{I_3}$ symmetry is preserved at the level of the Lagrangian.
This symmetry is broken when pions condense.

The mesonic potential consists of three pieces
\beq
V_M (\phi, \phi^*) 
 = V_{\rm sym} + V_{\rm SB} + V_{\rm anom} \,,
\eeq
with
a U(3) symmetric potential
\beq
V_{\rm sym} 
=
- \frac{\, m_M^2 \,}{\, 4 \,} \tr \big[ \calM \calM^\dag \big]
 +  \frac{\, \lambda \,}{\, 48 \,} \tr \big[ \big( \calM \calM^\dag \big)^2 \big] 
 \,,~~~~~~~
 \calM = \phi_a \lambda_a \,,
\eeq
and a symmetry breaking term associating the current quark masses, 
\beq
V_{\rm SB} 
=
 - \frac{\, c \,}{\, 2 \,} \tr \big[ \hat{m}_q^\dag \calM + \hat{m}_q \calM^\dag \big] \,,~~~~~~~~ \hat{m}_q = {\rm diag.} (m_l, m_l, m_s)
 \,,
\eeq
and the Kobayashi-Maskawa-'t Hooft (KMT) interaction 
\beq
 V_{\rm anom}
 =  - \frac{\, K \,}{2} \big[ \det \calM + \det \calM^\dag \big] \,.
\eeq
which is responsible for the U$_A(1)$ breaking.

%
%
%

Increasing $\mu_I$, the excitation energies of mesons including $u$ or $\bar{d}$ are reduced, i.e.,
the excitation energies of $\sigma^+, \pi^+, a_0^+, K^+, \cdots $ decrease linearly as a function of $\mu_I$.
As we see shortly, the tree level analyses show that the lightest pion condenses first,
and then the reduction of energies for the other mesons is stopped; no other mesons condense.
At tree level this is interpreted as the effective repulsion between condensed pions and the other mesons.
We assume this holds even after including loop corrections.
Thus we take the mean field as
\beq
\calM_{\rm MF} 
= 
\begin{bmatrix}
~  \la \sigma \ra & \rmi \la \pi_1 \ra & 0~\\
~  \rmi \la \pi_1\ra & \la \sigma \ra & 0~\\
~ 0 & 0 & \la \sigma_s \ra ~
\end{bmatrix}
\,,
~~~~~
\sigma = \sqrt{ \frac{2}{\, 3 \,} } \sigma_0 + \sqrt{ \frac{1}{\, 3 \,} }  \sigma_8 \,,
~~~~~
\sigma_s = \sqrt{ \frac{2}{\, 3 \,} } \sigma_0 - 2 \sqrt{ \frac{1}{\, 3 \,} } \sigma_8 \,.
\eeq
Here we could choose $\pi_1$ fields for condensed fields without loss of generality since the Lagrangian has the U(1)$_{I_3}$ symmetry.
The mean field Yukawa term now takes the form $(q_l=u,d)$
\begin{align}
\calL_{\rm Y}^{\rm MF}
= \bar{q}_l \big[ \rmi \Slash{\partial} + \mu_I \tau_3 \gamma_0 - M_l  - \rmi \gamma_5 \lambda_1 \Delta  \big] q_l
+   \bar{s} \big[ \rmi \Slash{\partial}  -  M_s  \big] s
 \,,
\end{align}
where we wrote $ M_l = g \la \sigma_l \ra$, $M_s = g \la \sigma_s \ra$, and $ \Delta = g \la \pi_1 \ra$.
Substituting the mean fields into the potentials, one finds
\beq
V_{\rm sym}^{\rm MF} 
=
- \frac{\, m_M^2 \,}{\, 2g^2 \,} ( M_l^2 + \Delta^2 ) 
 +  \frac{\, \lambda \,}{\, 24 g^4 \,} \big( M_l^2 + \Delta^2 \big)^2 
- \frac{\, m_M^2 \,}{\, 4g^2 \,} M_s^2  
 +  \frac{\, \lambda \,}{\, 48 g^4 \,} M_s^4
\eeq
and ($h_l = 2c m_l, h_s = 2 c m_s$)
\beq
V_{\rm SB}^{\rm MF} 
= -  \frac{\,  h_l \,}{g} M_l - \frac{\, h_s \,}{2g} M_s \,,
~~~~~~~~~~
V_{\rm anom}^{\rm MF} 
= -  \frac{\, K \,}{\, g^3 \,} \big( M_l^2 + \Delta^2 \big) M_s \,.
\eeq
Below we first examine the effective potential at tree level
which includes only the mesonic degrees of freedom.
The quark dynamics manifestly enters only after including quark loops.
As we see later, the quark loops change the structure of theories
and impose important constraints on the meson mean fields.

\section{Purely hadronic descriptions: tree level analyses}
\label{sec:tree}

\subsection{Gap equations at tree level}
\label{sec:gap_tree}

We first study the tree level potential including only meson degrees of freedom.
Combining the potentials with the terms including chemical potential in $\calL^M_{\rm kin}$,
the effective potential at tree level is
\beq
\Omega_0 
= - \frac{\, 2 \mu_I^2 \,}{g^2 } \Delta^2
+ V_{\rm sym}^{\rm MF} + V_{\rm SB}^{\rm MF} + V_{\rm anom}^{\rm MF} \,.
\eeq
The gap equations are
\begin{align}
\frac{\, \partial \Omega_0 \,}{\, \partial M_l \,}
&= 
\frac{\, M_l \,}{\, g^2 \,} 
 \bigg[ - m_M^2 
 + \frac{\, \lambda \,}{\, 6 g^2 \,} \big( M_l^2 + \Delta^2 \big) 
 - \frac{\, 2 K \,}{\, g \,} M_s
  -  \frac{\, g h_l \,}{\, M_l \,}
 \bigg]
 = 0 \,, 
 \notag \\
\frac{\, \partial \Omega_0 \,}{\, \partial M_s \,}
&= 
\frac{\, M_s \,}{\, 2g^2 \,} 
 \bigg[ - m_M^2 
 + \frac{\, \lambda \,}{\, 6 g^2 \,} M_s^2  
 - \frac{\, 2 K \,}{\, g \,} \frac{\, M_l^2 \,}{\, M_s \,}
  -  \frac{\, g h_s \,}{\, M_s \,}
 \bigg]
 = 0 \,,  
 \notag \\
 \frac{\, \partial \Omega_0 \,}{\, \partial \Delta \,}
&= 
\frac{\, \Delta \,}{\, g^2 \,} 
 \bigg[ - 4 \mu_I^2 
 - m_M^2 
 + \frac{\, \lambda \,}{\, 6 g^2 \,} \big( M_l^2 + \Delta^2 \big) 
 - \frac{\, 2 K \,}{\, g \,} M_s
 \bigg]
 = 0 \,.
\end{align}
We can show that $g h_l/M_l^{\rm vac} = m_\pi^2$.
We combine the first and third equations to derive a simple relation,
\beq
\frac{\, \partial \Omega_0 \,}{\, \partial \Delta \,}
= 
\frac{\, \Delta \,}{\, g^2 \,} 
 \bigg[ - 4 \mu_I^2 
 + \frac{\, g h_l \,}{\, M_l \,}
 \bigg]
 = 0 \,,~~~~~~~
 \frac{\, \partial^2 \Omega_0 \,}{\, (\partial \Delta)^2 \,}
= 
\frac{\, 1 \,}{\, g^2 \,} 
 \bigg[ - 4 \mu_I^2 
+ \frac{\, g h_l \,}{\, M_l \,}
+ \frac{\, \lambda \,}{\, 3 g^2 \,}  \Delta^2 
 \bigg] \,.
\eeq
The curvature with respect to $\Delta$ is positive for small $\mu_I$ so that $\Delta =0$ is favored.
Then the gap equation is the same as in the vacuum case;
in this domain $M_l = M_l^{\rm vac}$ and $g h_l/M_l^{\rm vac} = m_\pi^2$ holds.
The curvature becomes negative for $ m_\pi \ge 2 \mu_I $,  leading to the scaling $M_l \sim 1/\mu_I^2$.
This in turn leads to $\Delta^2 \sim 1/M_l \sim \mu_I^2$ from the condition $\partial \Omega_0/\partial M_l = 0$.
The strange quark is affected by isospin density only through the anomaly term.

\subsection{Meson kinetic and mass matrices at tree level}
\label{sec:meson}

We now quickly review the properties of mesons in a medium of condensed pions.
We first take a look at the excitation energies of charged mesons in the low density domain.
Then the Lagrangian has the following structure ($E = \sqrt{\vp^2 + m^2}$)
\beq
\frac{1}{\, 2 \,}
[ \phi_{+}, \phi_{-} ]
\begin{bmatrix}
~   0 ~ & ~ (p_0 + N_I \mu_I)^2 - E^2 ~\\
~   (p_0 + N_I \mu_I)^2 - E^2 & 0 ~
\end{bmatrix}
\begin{bmatrix}
 \phi_{+} \\
 \phi_{-}
\end{bmatrix}
\,,
\eeq
%
where $N_I =2$ for iso-triplet and $N_I = 1$ for iso-doublet.
The spectra are found to be
\beq
\omega_\pm (\vp) = E (\vp) \pm N_I \mu_I \,.~~~~~~({\rm in~normal~phase})
\eeq
Mesons with positive isospins have energy reduction.
This continues until one of those mesons condenses.
The pions are the lightest and have the largest isospin $N_I =2$ so they condense first, at $\mu_I = m_\pi/2$.

Before the pion condensation, a pair of positive and negative charged mesons together maintains the $\mu_I$-independence of the thermodynamics.
The thermodynamic potential from a (non-interacting) meson, for $N_I \mu_I \le E (\vp=0)$, is
\beq
\Omega
 = \frac{1}{\, 2 \,} \int_{\vp} \big( | E + N_I \mu_I | +   | E - N_I \mu_I | \big)
~\rightarrow~
 \frac{1}{\, 2 \,} \int_{\vp} \big( E + N_I \mu_I  + E - N_I \mu_I  \big) = \int_{\vp} E \,,
\eeq
after the $\mu_I$ terms cancel. The resulting energy is the zero point energy same as in the vacuum.
The expression is valid until $N_I \mu_I$ reaches the meson mass.

After pions condense, mesons having energy reduction in the normal phase are now subject to effective repulsions
and hence the gap remains in the excitation energies.
In addition condensed pions supply isospin and parity violating sources so that various mesons mix.
At tree level the mixing is caused through the quartic and the KMT interaction.
We read off the meson mass matrices masses by looking at the quadratic order of the potential
\begin{align}
&\hspace{-0.5cm}
V_M^{\rm quad} 
=  
  \frac{\, \partial^2 \tilde{V}_0 \,}{\, \partial \phi_a \partial \phi^*_b \,} \phi_a \phi_b^*
+ \frac{1}{\, 2 \,} \frac{\, \partial^2 \tilde{V}_0 \,}{\, \partial \phi_a \partial \phi_b \,}  \phi_a \phi_b
+ \frac{1}{\, 2 \,} \frac{\, \partial^2 \tilde{V}_0 \,}{\, \partial \phi_a^* \partial \phi_b^* \,}  \phi_a^* \phi_b^*
+ \cdots
\notag \\
& \hspace{-0.5cm}
=
 \frac{\, \partial^2 \tilde{V}_0 \,}{\, \partial \phi_a \partial \phi^*_b \,}  \big[ \sigma_a \sigma_b + \pi_a \pi_b  \big] 
	+ {\rm Re} \frac{\, \partial^2 \tilde{V}_0 \,}{\, \partial \phi_a \partial \phi_b \,} \big( \sigma_a \sigma_b - \pi_a \pi_b \big)  
	-  {\rm Im} \frac{\, \partial^2 \tilde{V}_0 \,}{\, \partial \phi_a \partial \phi_b \,} \big( \sigma_a \pi_b + \pi_a \sigma_b \big)  
\end{align}
where we evaluated the second derivative at the mean field values,
and also used the fact that  $\frac{\, \partial^2 \tilde{V}_0 \,}{\, \partial \phi_a \partial \phi^*_b \,} \big|_{ {\rm MF} }$ is real.
The mass matrices for $\sigma$, $\pi$ are
\beq
 m_{\sigma_a \sigma_b}^2 
= 2\bigg[ \frac{\, \partial^2 \tilde{V}_0 \,}{\, \partial \phi_a \partial \phi^*_b \,}
+ {\rm Re} \frac{\, \partial^2 \tilde{V}_0 \,}{\, \partial \phi_a \partial \phi_b \,}
\bigg]
\,,~~~~
m_{\pi_a \pi_b}^2 
= 2\bigg[ \frac{\, \partial^2 \tilde{V}_0 \,}{\, \partial \phi_a \partial \phi^*_b \,}
- {\rm Re} \frac{\, \partial^2 \tilde{V}_0 \,}{\, \partial \phi_a \partial \phi_b \,}
\bigg] \,,
\eeq
and the parity breaking $\sigma$-$\pi$ couplings induced by the pion condensate are 
\beq
m_{\sigma_a \pi_b}^2 = - 2 {\rm Im} \frac{\, \partial^2 \tilde{V}_0 \,}{\, \partial \phi_a \partial \phi_b \,}~~~(\propto \Delta ) \,.
\eeq
The pion condensate $\pi_1 \sim u \bar{d} + d \bar{u}  $ induces the conversion $ u \leftrightarrow d $ and $\bar{u} \leftrightarrow \bar{d} $,
and also causes the conversion between the $\sigma$- and $\pi$-sectors.
For light flavors all channels but $\pi_3, \sigma_3 \sim u \bar{u} - d \bar{d} $ mix.
The exceptional $\pi_3$ and $\sigma_3$ are protected from the mixing because of the G-parity \cite{Lee:1956sw} type symmetries, as we discuss shortly.
All kaons are mixed one another. 
But the iso-doublets do not mix with iso-scalar nor iso-vector since the condensed $\pi_1$ supply integer isospins.

To understand the decoupling of $\pi_3$ and $\sigma_3$ from the others, we consider unitary transformations induced by
\beq
P_{I_3} (\theta) = P \rme^{\rmi \theta I_3} \,,~~~~~~~
C_{I_1} (\theta) = C \rme^{\rmi \theta I_1} \,, 
\eeq
where $P$ and $C$ are parity and charge conjugation, respectively.
We define $ C^{-1} \pi_+ C = \pi_- $ and $C^{-1} \pi_3 C = \pi_3$ so that the G-parity is $-1$ for all pions.
In this definition $ C^{-1} \pi_1 C = \pi_1 $ and $ C^{-1} \pi_2 C = -\pi_2 $.

The $P_{I_3}$ symmetry holds in our U$_{I_3}$ symmetric Lagrangian for any $\theta$.
The question is whether this symmetry is spontaneously broken or not.
The $\pi_1$ condensate under this symmetry transforms as
\beq
P_{I_3}^{-1} (\theta) \pi_1 P_{I_3} (\theta)
\sim  P^{-1}  \big( \rme^{-\rmi \theta } \bar{u} \rmi \gamma_5 d + \rme^{\rmi \theta} \bar{d} \rmi \gamma_5 u \big)  P
= - \big( \rme^{ - \rmi \theta } \bar{u} \rmi \gamma_5 d + \rme^{ \rmi \theta} \bar{d} \rmi \gamma_5 u \big)  \,,
\eeq
which in general is the mixture of $\pi_1$ and $\pi_2$.
But setting $\theta = \pi$ and writing $\calP_{I_3} \equiv P_{I_3}(\pi)$, we conclude that
$ \calP_{I_3}^{-1}  \pi_1 \calP_{I_3}  = \pi_1$.
Thus the $\calP_{I_3}$-parity is conserved in the presence of the $\pi_1$ condensates.
We also note that $\calP_{I_3}$-parity is also not violated by $\sigma$ and $\sigma_s$ condensates.
Thus $\calP_{I_3}$ can be used to classify various mesons.
Under the $\calP_{I_3}$ transformation, the isoscalar and isovectors transform as
\beq
\calP_{I_3}^{-1} \big( \sigma_{0,3,8}, \pi_{1,2} \big) \calP_{I_3} = + \big( \sigma_{0,3,8}, \pi_{1,2} \big) \,,~~~~~
\calP_{I_3}^{-1} \big( \sigma_{1,2}, \pi_{0,3,8} \big) \calP_{I_3} = - \big( \sigma_{1,2}, \pi_{0,3,8} \big)  \,.
\eeq
Next we consider $C_{I_1} (\theta)$.
First we note the isospin density is invariant for $\calC_{I_1} \equiv C_{I_1} (\pi)$,
\beq
C_{I_1}^{-1} (\theta) \big( \bar{u} u - \bar{d} d \big) C_{I_1} (\theta)
= - C^{-1} \big( \bar{u} u - \bar{d} d \big) C = \bar{u} u - \bar{d} d \,,
\eeq
where we used $\rme^{ - \rmi \pi I_1} (u,d) \rme^{\rmi \pi I_1} = (d,u)$. 
The Lagrangian is also invariant.
The condensates $\pi_1$, $\sigma$ and $\sigma_s$ are also invariant.
Under the $\calC_{I_1}$ transformation, the isoscalar and isovectors transform as
\beq
\calC_{I_1}^{-1} \phi_{0, 1,2, 8} \calC_{I_1} = + \phi_{0,1,2,8} \,,~~~~~
\calC_{I_1}^{-1} \phi_3 \calC_{I_3} = - \phi_3 \,.
\eeq
Combining these symmetries, the isoscalar and isovector sectors can be decomposed into
\beq
\sigma_3\,, ~~~~~~\pi_3 \,, ~~~~~~ (\sigma_0, \sigma_8 ; \pi_1, \pi_2)\,,~~~~~~~(\pi_0, \pi_8 ; \sigma_1, \sigma_2)\,.
\eeq
Meanwhile, for the isodoublet sectors
\beq
\big( \sigma_4\,, \sigma_5\,, \sigma_6\,, \sigma_7\, ;  \pi_4\,, \pi_5\,, \pi_6\,, \pi_7\, \big)\,
\eeq
are all mixed.

\subsubsection{The spectra of $\sigma_3$, $\pi_3$,  and $(\sigma_0, \sigma_8 ; \pi_1, \pi_2)$.}
\label{sec:massless}

Here we display some analytic expressions for $\sigma_3$, $\pi_3$,  and $(\sigma_0, \sigma_8 ; \pi_1, \pi_2)$.
The $\sigma_3$ and $\pi_3$ decouple from the other modes and the mass matrices are
\begin{align}
 m^2_{\pi_3}
 &
 = 
 - m_M^2 
 +  \frac{\, \lambda \,}{\, 6 g^2 \,} (M_l^2 + \Delta^2)  
 - \frac{\, 2K \,}{\, g \,} M_s 
 = 4 \mu_I^2 \,,
 \notag \\
 m^2_{\sigma_3}
 &
 = 
 - m_M^2 
 + \frac{\, \lambda \,}{\, 2 g^2 \,} (M_l^2 + \Delta^2)  
 + \frac{\, 2K \,}{\, g \,} M_s
= 4 \mu_I^2 
 + \frac{\, \lambda \,}{\, 3 g^2 \,} (M_l^2 + \Delta^2)  
 + \frac{\, 4K \,}{\, g \,} M_s \,,
 \end{align}
 where for $\pi_3$ we have used the gap equation.
 For these neutral modes the excitation energies directly coincide with the mass,
 $\omega_{\pi_3, \sigma_3} = m_{\pi_3, \sigma_3}$.
 
 Next we look into the charged sector. The pion mass matrices are
\begin{align}
m^2_{\pi_1}
 &
 = 
 - m_M^2 
 +  \frac{\, \lambda \,}{\, 6 g^2 \,} (M_l^2 + 3 \Delta^2)  
 - \frac{\, 2K \,}{\, g \,} M_s
 =
 4\mu_I^2 +  \frac{\, \lambda \,}{\, 3 g^2 \,} \Delta^2  
 \,,
 \notag \\
 m^2_{\pi_2}
 &
 = 
 - m_M^2 
 +  \frac{\, \lambda \,}{\, 6 g^2 \,} (M_l^2 + \Delta^2)  
 - \frac{\, 2K \,}{\, g \,} M_s 
 = 4 \mu_I^2 \,,
 \end{align}
 and the neutral $\sigma$-sector has the mass matrices
 \begin{align}
 m^2_{\sigma_0}
&=
 - m_M^2 
 +  \frac{\, \lambda \,}{\, 18 g^2 \,} \big[ 6 M_l^2 + 3M_s^2 + 2 \Delta^2 \big] 	
 - \frac{\, 4K \,}{\, 3g \,}  \big[  2M_l + M_s \big] 
 \,,
\notag \\
m^2_{\sigma_8}
&=
 - m_M^2 
 +  \frac{\, \lambda \,}{\, 18 g^2 \,} \big[ 3 M_l^2 + 6 M_s^2 + 2 \Delta^2 \big] 	
 + \frac{\, 2K \,}{\, 3g \,}  \big[  4 M_l - M_s \big] 
 \,.
\end{align}
The mixing is induced by
\begin{align}
m_{\sigma_0 \pi_1}^2 
 = 
  \frac{\, \sqrt{2} K \,}{\, \sqrt{3} g \,}  \Delta 
  \,,
~~~~~~~
m^2_{\sigma_0 \sigma_8}
=
 \frac{\, \lambda \sqrt{2} \,}{\, 12 g^2 \,}  \Delta^2 
- \frac{\, \lambda \sqrt{2} \,}{\, 36 g^2 \,} \delta M_s^2 
+ \frac{\, \sqrt{2} K \,}{\, 3g \,} \delta M_s 
 \,.
 \end{align}
 The coupling between the $\sigma_{0,8}$ and $\pi_{1,2}$ sectors is induced through the convolution of U$_A(1)$ anomaly and the pion condensate.
 Meanwhile, the mixing between $\sigma_0$ and $\sigma_8$ is due to the explicit SU(3) breaking
 which appears through $\delta M_s = M_s - M_l$ and the meson condensates existing only in the $u,d$ sector.

The isospin chemical potentials in the kinetic terms and mass terms together yield the massless modes.
For a simple illustration, we consider the case where $K$ is small.
In this approximation $\pi_{1,2}$ and $\sigma_{0,8}$ decouple.
Now the terms quadratic in $\pi_{1,2}$ are
\beq
\begin{bmatrix}
~   p_0^2 - m_{\pi_1}^2 + 4 \mu_I^2 ~&~ 4 \rmi \mu_I p_0 ~\\
~  - 4 \rmi \mu_I p_0 ~&~ p_0^2 - m_{\pi_2}^2 + 4 \mu_I^2 ~
\end{bmatrix}
 =
\begin{bmatrix}
~   p_0^2 -  \frac{\, \lambda \,}{\, 3 g^2 \,} \Delta^2 ~&~ 4 \rmi \mu_I p_0 ~\\
~  - 4 \rmi \mu_I p_0 ~&~ p_0^2  ~
\end{bmatrix}
\,.
\eeq
We have used the expression of $m^2_{\pi_1}$ and $m^2_{\pi_2}$.
The determinant of the matrix becomes zero at $p_0 =0$, reflecting that there is a massless mode
associated with the spontaneous breakdown of the U$_{I_3}$ symmetry.

\subsubsection{The spectra of $(\sigma_{4-7} ; \pi_{4-7} )$.}
\label{sec:kaons}

Next we discuss the isodoublet (kaon) sector.
The diagonal part is
\begin{align}
m^2_{\pi_{4-7} }
 &
 = 
 - m_M^2 
 +  \frac{\, \lambda \,}{\, 6 g^2 \,} (M_l^2 + M_s^2 - M_l M_s + \Delta^2  )  
 - \frac{\, 2K \,}{\, g \,} M_l
 \notag 
 \\
 &= 4\mu_I^2 
+  \frac{\, \lambda \,}{\, 6 g^2 \,} M_l ( M_s - M_l )
+ \frac{\, 2K \,}{\, g \,}  ( M_s - M_l ) \,,
 \notag \\
  m^2_{\sigma_{4-7} }
 &
 = 
 - m_M^2 
 +  \frac{\, \lambda \,}{\, 6 g^2 \,} (M_l^2 +M_s^2 + M_l M_s + \Delta^2 )  
 + \frac{\, 2K \,}{\, g \,} M_l
  \notag 
 \\
 &= 4\mu_I^2 
+  \frac{\, \lambda \,}{\, 6 g^2 \,} M_s ( M_s + M_l )   
+ \frac{\, 2K\,}{\, g \,}  ( M_s + M_l )
 \,,
\end{align}
where we have used the gap equation for the pion sector.
The difference in mass matrices between the pion and kaon sectors come from the effective quark mass.
The mixing matrices are
\begin{align}
& m^2_{\sigma_4, \sigma_7} = - m^2_{\sigma_5, \sigma_6} 
= - m^2_{\pi_4, \pi_7} = m^2_{\pi_5, \pi_6} 
 =  \frac{\, \lambda \,}{\, 24 g^2 \,} \Delta \delta M_s
   \,,
   \notag \\
&  m_{\sigma_4 \pi_6}^2 
= m_{\sigma_5 \pi_7}^2 
= m_{\pi_4 \sigma_6}^2 = m_{\pi_5 \sigma_7}^2 
=  \frac{\, \lambda \,}{\, 24 g^2 \,} \Delta  \big( M_l + M_s  \big)
 - \frac{\, K \,}{\, g \,}  \Delta   
 \,.
\end{align}
These form a $8 \times 8$ matrix and need numerical analyses for the determination of the spectra.

To get analytic insights we consider the case where $M_l, M_s \ll \mu_I$ for which the mixing is neglected compared to the diagonal part.
Combining the mass matrices with the kinetic term, (we write $m_K^2 = m_{4-7}^2$)
\beq
\begin{bmatrix}
~   p_0^2 - m_{K}^2 + \mu_I^2 ~&~ 2 \rmi \mu_I p_0 ~\\
~  - 2 \rmi \mu_I p_0 ~&~ p_0^2 - m_{K}^2 +  \mu_I^2 ~
\end{bmatrix}
 \sim
\begin{bmatrix}
~   p_0^2 - 3 \mu_I^2 ~&~ 2 \rmi \mu_I p_0 ~\\
~  - 2 \rmi \mu_I p_0 ~&~ p_0^2 - 3 \mu_I^2  ~
\end{bmatrix}
\,.
\eeq
For kaons being the isodoublet, the energy reduction associated with the chemical potential is weaker than in the pion,
and hence $\mu_I$ terms do not cancel. 
The excitation energies manifestly depend on $\mu_I$,
\beq
\omega_K ~\sim~ \mu_I \,,~~ 3\mu_I \,.
\eeq
The same also holds for the scalar meson $\kappa$.

\section{Quark descriptions}
\label{sec:quark_loop}

Now we include loops made of mean-field quark propagators.
We keep only one loop with the leading $\Nc$ contributions;
quark loops for the vertex corrections are neglected.
In counting of loops the quark contributions appear as radiative ``corrections'' to the tree level,
but in fact the quark contributions should be regarded as leading order contributions at finite density \cite{Chiba:2023ftg};
indeed as $\mu_I$ increases the quark contributions dominate over hadronic contributions
and change the structure of gap equations and EOS.

\subsection{The structure of the effective potential}
\label{sec:eff_pot}

Beyond tree calculations, the parameters in the effective potential must be also reinterpreted as bare parameters
which are split into the renormalized parameters and counter terms.
First we attach an index $B$ to fields $\phi$, $q$, $m_M$, and so on, and factor out the renormalized parameters and fields  \cite{Chiba:2023ftg},
\beq
\phi_B &=& Z_{\phi}^{1/2} \phi\,, ~~~~~~ q_B = Z_{q}^{1/2}  q \,, \notag \\
g_B &=& \tilde{Z}_g Z_q^{-1} Z_{\phi}^{-1/2} g = Z_g g\,, \notag \\
(m_{M}^2)_B &=& \tilde{Z}_{m^2} Z_\phi^{-1} m_M^2 =  Z_{m^2} m_M^2\,, \notag \\
\lambda_{ B} &=& \tilde{Z}_\lambda Z_\phi^{-2} \lambda = Z_\lambda  \lambda\,, \notag \\
(h_l)_B &=& \tilde{Z}_{h_l} Z_\phi^{-1/2} h = Z_{h_l} h_l \,,\notag \\
(h_s)_B &=& \tilde{Z}_{h_s} Z_\phi^{-1/2} h = Z_{h_s} h_s \,.
\label{eq:reno_para}
\eeq
We also use $\delta Z_i = Z_i - 1$ with $i=\phi, q, g$, and so on.
The $\tilde{Z}_i$ represents the radiative corrections without those for the external lines.
The loop corrections to the quark self-energies and quark-meson vertices
appear only through meson-loops and hence
\beq
 Z_\psi = 1 + O(1/\Nc)\,,~~~ \tilde{Z}_g = 1 + O(1/\Nc)\,.
\eeq
Meanwhile, the meson self-energies and tadpole contain quark loops of $O(\Nc)$
which are combined with $g^2 \sim 1/\Nc$ vertices to yield
\beq
&& Z_\phi = 1 + O( g^2 \Nc ) \,,~~~ \tilde{Z}_{m^2} = 1 + O( g^2 \Nc ) \,, \notag \\
&& \tilde{Z}_h  = 1 + O(g^2\Nc) \,,~~~ \tilde{Z}_\lambda = 1 + O(g^4\Nc/\lambda)\,.
\eeq
and hence one must keep these corrections.
It is useful to note that the relation
\beq
g_B \phi_B = g \phi \,,
~~~~~
Z_g = Z_\phi^{-1/2} \,,
\label{eq:g_and_Z}
\eeq
in the large $\Nc$ limit.
With the first relation the dynamically generated quark mass and gap are RG invariant.
The second relation tells that  the running of $g^2$ can be studied by examining the meson propagators.

The effective potential up to one-loop consists of three type of terms
\beq
\Omega = \Omega_0 + \Omega_{\rm c.t.} + \Omega_q \,.
\eeq
The $\Omega_0$ takes the same form as in the tree level but parameters are to be interpreted as renormalized ones.
The $\Omega_{\rm c.t.}$ includes the counter terms originating from the splitting, e.g., $\lambda_B = \lambda + \delta \lambda$,
\beq
\Omega_{c.t.} 
&= - \delta Z_\phi \frac{\, 2 \mu_I^2 \,}{g^2 } \Delta^2
- \frac{\, \delta m_M^2 \,}{\, 2g^2 \,} ( M_l^2 + \Delta^2 ) 
 +  \frac{\, \delta \lambda \,}{\, 24 g^4 \,} \big( M_l^2 + \Delta^2 \big)^2 
  -  \frac{\, \delta h_l \,}{g} M_l  
  \notag \\
&
- \frac{\, \delta m_M^2 \,}{\, 4g^2 \,} M_s^2  
 +  \frac{\, \delta \lambda \,}{\, 48 g^4 \,} M_s^4
 - \frac{\, \delta h_s \,}{2g} M_s
 -  \frac{\, \delta K \,}{\, g^3 \,} \big( M_l^2 + \Delta^2 \big) M_s \,.
\eeq
The quark contribution is
\beq
\Omega_q =  
 - 2\Nc \int_{\vp} \big( 
 \xi_+  + \xi_-  + E_s 
  \big) \,,
\eeq
%
where $E_l = \sqrt{ \vp^2 + M_l^2}$ and $E_s = \sqrt{ \vp^2 + M_s^2 }$,
and
\beq
\xi_- = \xi_u = \xi_{\bar{d}} = \sqrt{ \big( E_l - \mu_I \big)^2+\Delta^2 } \,,~~~~~~~
\xi_+ = \xi_d = \xi_{\bar{u}} = \sqrt{ \big( E_l + \mu_I \big)^2+\Delta^2 } \,.
\eeq
In the $u,d$-quark sector quarks have the energy dispersion of the BCS type.

For small $\mu_I \le m_\pi/2 \le M_l$, the pion condensate is absent.
Then the chemical potential disappears from the $\Omega_q$,
\beq
\sqrt{ \big( E_l + \mu_I \big)^2+\Delta^2 } 
  +\sqrt{\big( E_l - \mu_I \big)^2+\Delta^2} 
  ~ \rightarrow ~ 
   (E_l + \mu_I) + (E_l -\mu_I) = 2E_l \,,
   \label{eq:bcs_gap}
\eeq
reflecting the Sliver Blaze property that the isospin density $n_I $
begins to change only when either $\mu_I \ge M_l$ or $\Delta \neq 0$ are satisfied.
In the present case $\Delta \neq 0$ is the driving force.

The $\Omega_q$ is UV divergent.
This divergence is cancelled by the counter terms from the masses, couplings, and field normalization factor from the mesonic Lagrangian.
We isolate the divergences as \cite{Andersen:2018qkq}
\beq
\Omega_q 
= \Omega_l^{\rm R} 
 -2 \Nc \int_{\vp}
 \bigg[ 2 \sqrt{E_q^2+\Delta^2}
  + \frac{\, \mu_I^2\Delta^2 \,}{\, (E_q^2 + \Delta^2 \,)^{3/2}} 
  + E_s
  \bigg] 
 \eeq
 where the twice subtracted energy density, including only $u,d$-quark contributions, is
\beq
 \Omega_l^{\rm R}
 =-2\Nc
\int_p \bigg[
  \xi_+ + \xi_-
  - 2 \sqrt{E_q^2+\Delta^2}
  - \frac{\, \mu_I^2\Delta^2 \,}{\, (E_q^2 + \Delta^2 \,)^{3/2}} \bigg] 
  \,.
\eeq
The $\Omega_l^R$ is UV finite and vanishing when $\mu_I =0$ and $\Delta =0$.
At large density, the subtracted energy scales as $\Omega_l^R \simeq \mu_I^4$, dominating the thermodynamic potential.

To understand the physical meaning of the subtracted potential $\Omega_l^R$ at large $\mu_I$,
it is instructive to consider $\Delta \rightarrow 0$ limit and decompose the integral into the $E_l \le \mu_I$ and $E_l \ge \mu_I$ domains.
For $E_l \le \mu_I$,
\beq
 \int_{\vp} \Theta  (\mu_I - E_l ) \big(\, | E_l + \mu_I | + | E_l - \mu_I | - 2E_l \, \big)
= 2 \int_{\vp} \Theta ( \mu_I - E_l ) \, \big( \mu_I - E_l \big) \,,
\eeq
and for $ E_I \ge \mu_l $,
\beq
 \int_{\vp} \Theta  ( E_l - \mu_I ) \big(\, | E_l + \mu_I | + | E_l - \mu_I | - 2E_l \, \big) = 0  \,.
\eeq
Thus, $\Omega_l^R$ at $\Delta =0$ is nothing but the standard quasi-particle contributions to the thermodynamic potential.

The divergent piece in $\Omega_q$ is evaluated in the $\overline{MS}$ scheme,
\begin{align}
\Omega_q^{\rm UV} 
&
 = \frac{\, 2 \Nc  \,}{\, (4\pi)^2 \,} (M_l^2 + \Delta^2)^2 \bigg[ \frac{1}{\, \epsilon \,} + \frac{\, 3 \,}{2} + \ln \frac{\, \Lambda^2 \,}{\, M_l^2 + \Delta^2 \,} \bigg]
 - \frac{\, 8 \Nc  \,}{\,  (4\pi)^2 \,} \mu_I^2 \Delta^2 \bigg[ \frac{1}{\, \epsilon \,} + \ln \frac{\, \Lambda^2 \,}{\, M_l^2 + \Delta^2 \,} \bigg]
 \notag \\
&~~~
 + \frac{\,  \Nc  \,}{\, (4\pi)^2 \,} M_s^4 \bigg[ \frac{1}{\, \epsilon \,} + \frac{\, 3 \,}{2} + \ln \frac{\, \Lambda^2 \,}{\, M_s^2 \,} \bigg]
 \,,
\end{align}
where $\Lambda$ is the renormalization scale.
In the $\overline{MS}$ scheme, the counter terms are used to absorb only the $1/\epsilon$ terms.
Then the counter terms are fixed as
\beq
\delta m_M^2 = 0\,,~~~~~~
\delta{h}_{l,s} = 0 \,,~~~~~~ 
\delta \lambda = - \frac{\, 48 \Nc g^4 \,}{\, (4\pi)^2 \epsilon \,} \,,~~~~~~
\delta Z_\phi = - \frac{\, 4 \Nc g^2 \,}{\, (4\pi)^2 \epsilon \,} \,,~~~~~~
\delta K = 0 \,.
\eeq
In our previous studies we further discussed the on-mass-shell renormalization
to tune the pole positions of $\pi$ and $\sigma$.
We do this when we fix parameters of our model.

Putting everything together, the effective potential $\Omega = \Omega_0 + \Omega_{\rm c.t.} + \Omega_q$ are reorganized as
\beq
\Omega = \Omega_l + \Omega_s + \Omega_{\rm mix}
\eeq
where the contributions from $u,d$-quarks are
\begin{align}
\Omega_l
&= \Omega_l^R 
- \frac{\, m_M^2 \,}{\, 2g^2 \,} ( M_l^2 + \Delta^2 ) 
 +  \frac{\, \lambda \,}{\, 24 g^4 \,} \big( M_l^2 + \Delta^2 \big)^2 
 -  \frac{\, h_l \,}{g} M_l 
  \notag \\
&
+  \frac{\, 2 \Nc  \,}{\, (4\pi)^2 \,} (M_l^2 + \Delta^2)^2 \bigg[  \frac{\, 3 \,}{2} - \ln \frac{\, M_l^2 + \Delta^2 \,}{\, M_0^2 \,} \bigg]
 + 2 \mu_I^2 \Delta^2 \bigg[ -\frac{\, 1 \,}{g^2} + \frac{\, 4\Nc \,}{\, (4\pi)^2 \,} \ln \frac{\, M_l^2 + \Delta^2 \,}{\, M_0^2 \,} \bigg]
\end{align}
and from $s$-quarks
\beq
\Omega_s
= \frac{1}{\, 2 \,} \bigg( 
  - \frac{\, m_M^2 \,}{\, 2g^2 \,} M_s^2  
 +  \frac{\, \lambda \,}{\, 24 g^4 \,} M_s^4 
	 -  \frac{\, h_s \,}{g} M_s 
	+  \frac{\, 2 \Nc  \,}{\, (4\pi)^2 \,} M_s^4 \bigg[  \frac{\, 3 \,}{2} - \ln \frac{\, M_s^2 \,}{\, M_0^2 \,} \bigg]
\bigg)
\eeq
and the $(u,d)$-$s$ quark mixing induced by the KMT interaction,
\beq
\Omega_{\rm mix} 
= -  \frac{\, K \,}{\, g^3 \,} \big( M_l^2 + \Delta^2 \big) M_s \,.
\eeq
We note that, the strange quark sector shows the substantial $\mu_I$ dependence because of the U$_A$(1) induced coupling, $\sim \Delta^2 M_s$;
the $M_s$ increases as $\Delta$ does.
The similar effects have been studied in the context of color-superconductivity \cite{Hatsuda:2006ps,Yamamoto:2007ah,Hatsuda:2008is,Zhang:2008ima}.

The expression of $\Omega$ is the renormalization group (RG) invariant for a given field values, 
provided that the renormalized masses, couplings, etc. follow the RG evolution.
We have set the renormalization scale $\Lambda$ to be the vacuum light quark mass, $M_0 \equiv M_l^{\rm vac}$,
and then the couplings etc should be interpreted as those evaluated at $\Lambda = M_0$.

Finally we mention that the expression is invalid for very large $g^2 \ln \big( M^2+\Delta^2 \big) $ \cite{Coleman:1973jx}; as is well-known, the one-loop effective potential is unbound for large field limits;
there we need higher orders in loops.

\subsection{The structure of the gap equations}
\label{sec:gap_quark}

The quark loops have added the logarithmic terms which change the structure of the effective potential, especially at large $\mu_I$.
Let us see this assuming $M_l, M_s \ll \mu_I$.
At tree level the effective potential behaves as
\beq
\Omega_{\rm tree} 
~\rightarrow~
- \frac{\, 2 \mu_I^2\,}{\, g^2 \,} \Delta^2  
 +  \frac{\, \lambda \,}{\, 24 g^4 \,} \Delta^4 \,.
\eeq
The gap equation for $\Delta$ balances the first and second terms, so $\Delta_{\rm tree} \sim \mu_I$ inevitably follows.
Substituting back the solution $\Delta_{\rm tree}^2 \sim 24 \mu_I^2 g^2/\lambda $ into the $\Omega_{\rm tree}$,
we find
\beq
\Omega_{\rm tree} 
~ \sim ~
- \frac{\, 24 \,}{\, \lambda \,} \mu_I^4 \,.
\eeq
The hadronic parameter $\lambda$ entirely fixes the coefficient of $\mu_I^4$ term.

This trend changes after including quark contributions. 
We assume that the gap equation leads to $M_{l,s}, \Delta \ll \mu_I$, and then check the consistency of this assumption.
The large $\mu_I$ behavior of $\Omega$ is
\beq
\Omega 
~\rightarrow~
\Omega_l^R 
+ 2 \mu_I^2 \Delta^2 \bigg[ - \frac{\, 1 \,}{g^2} + \frac{\, 4\Nc \,}{\, (4\pi)^2 \,} \ln \frac{\, \Delta^2 \,}{\, M_0^2 \,} \bigg] \,,
\eeq
where the tree level terms, except $\mu_I^2/g^2$ term, are suppressed compared to $\mu_I^2$ and $\mu_I^4$ terms.
We note that the logarithmic contribution changes the sign at $M_0$;
it becomes attractive contributions for $\Delta \lesssim M_0$ but repulsive for $\Delta \gtrsim M_0$.
At large density the latter stops the growth of $\Delta$ induced by the tree contribution before $\lambda \Delta^2$ terms are activated.

It turns out that $\Omega_l^R$ is well saturated by $\mu_I^4$ terms and weakly depend on $\Delta$.
Then the gap equation is dominated by the coefficients of $\mu_I^2$ term as
\beq
\frac{\, \partial \Omega \,}{\, \partial \Delta^2 \,} 
~\rightarrow~
 2\mu_I^2\bigg[ 
 - \frac{1}{\, g^2 \,} + \frac{\, 4\Nc \,}{\, (4\pi)^2 \,} + \frac{\, 4\Nc \,}{\, (4\pi)^2 \,} \ln \frac{\, \Delta^2 \,}{\, M_0^2 \,} 
 \bigg]
 + O(M_0^2)
~\simeq~ 0 \,.
\eeq
This form with the logarithm is typical for low dimensional systems; in the present case the Fermi surface plays the role for the dimensional reduction.
Unlike the tree level solution, at $M_0/\mu_I \rightarrow 0$
the gap does not manifestly contain the $\mu_I$-dependence, 
\beq
\Delta^2 \simeq M_0^2 \, \rme^{ - 1 + \frac{\, 4\pi^2 \,}{\, \Nc g^2 \,}  } \,.
\eeq
The $g$ dependence is a bit puzzling; 
the usual BCS type calculations have the form of $\rme^{- \#/G}$ with a coupling constant $G$,
meaning that a stronger $G$ increases the size of the gap.
In contrast, a large $g$ reduces the gap in our model.
Our interpretation is the following:
the condensed mesons at tree level, where quarks inside are neglected,
overreact to the increase of $\mu_I$ and get overpopulated, as one can see from the scaling $\Delta_{\rm tree} \sim \mu_I$.
In particular the quark Pauli blocking is neglected.
In light of this viewpoint, the explicit inclusion of quarks with the coupling to mesons 
should cancel the unwanted contributions and should temper the growth of $\Delta$. 
This effect is stronger for a larger $g$.
For a very small $g$, apparently $\Delta$ blows up, but then $\lambda \Delta^4$ terms become dominant and the situation goes back to the tree level model or purely mesonic models.
The philosophy here is in line with Ref.~\cite{Duarte:2021tsx} that introduced ghosts to cancel double counted contributions.

Substituting back the $\Delta$ into $\Omega$, the thermodynamic potential at very large $\mu_I$ scales as
\beq
\Omega 
~ \sim ~ 
\Omega_l^R 
 - \frac{\, \Nc \,}{\, 2 \pi^2 \,} \mu_I^2 \Delta^2
   \,,
\eeq
Note that, in $\Omega$ at large $\mu_I$, 
the mesonic parameters $m_M^2$ and $\lambda$ no longer play important roles, while the Yukawa coupling $g$ is now hidden into the gap parameter $\Delta$.

\begin{figure}[t]
\vspace{-.0cm}
\begin{center}
\includegraphics[width=9. cm]{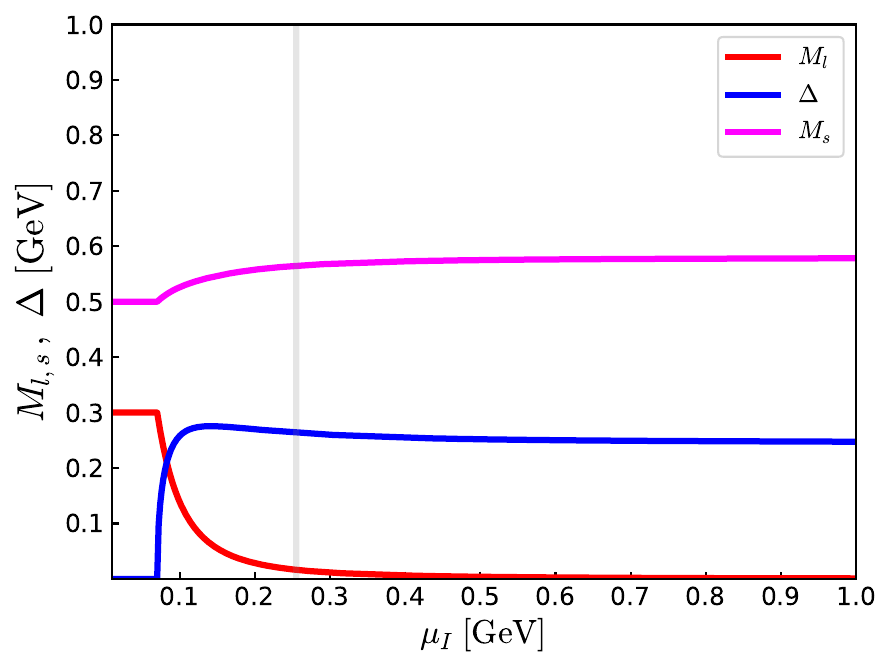}
\end{center}
\vspace{-0.0cm}
\caption{ The effective quark masses $M_l, M_s$ and the BCS gap $\Delta$ as functions of $\mu_I$ in the quark-meson model with the set B in Table.~\ref{tab:coupling}.
The growth of the strange quark mass comes from the U$_A$(1) anomaly through which strange quarks couple to the pion condensate.
}
\label{fig:gap}
\end{figure}   

\section{ Meson poles at one-loop: parameter fixing }
\label{sec:meson_poles}

Now we have all expressions needed to discuss EOS
but still the parameters in our model remain to be fixed.
We use meson nonets in vacuum to fix our parameters.
Here the meson self-energies induced by quark loops 
must be taken into account for the consistency with the gap equations,
Meanwhile in this work we do not address in-medium self-energies to meson masses.

\subsection{Meson poles }
\label{sec:meson}
As we have constructed the effective potential at one-loop,
we need to upgrade the tree level expressions of meson masses.
This is essential to reproduce the masses of the Goldstone bosons which arise at the minima of the effective potential at a given order of approximation.
The analyses go as in the Nambu-Jona-Lasinio model \cite{Hatsuda:1994pi}.

An inverse of meson propagator with loops takes the form
\beq
-p^2 + m^2 = - p^2 + m^2_{\rm tree} + \Sigma^q (m^2) + \Sigma^{c.t.} (m^2) \,,
\eeq
where $\Sigma^q$ comes from quark loops while $\Sigma^{c.t.}$ from the counter terms.
In case of pions, we have
\beq
- p^2 + m_\pi^2 
= \bigg( - m_M^2 
 + \frac{\, \lambda \,}{\, 2 g^2 \,} M_0^2
 - 2 \frac{\, K \,}{\, g \,} M_s 
 \bigg)
 + \Sigma_\pi^q (m_\pi^2)
- p^2 \delta Z_\phi  + \frac{\, \delta \lambda \,}{\, 2 g^2 \,} M_0^2 \,.
\eeq
We obtained the counter terms from the tree level bare parameters.

It is convenient to define
\begin{align}
&\Sigma_{fg}^{q} (p^2)
= \rmi  g^2 \int_l \tr_{\rm c,D} \big[ S_f (l) \rmi \gamma_5 S_g (p+l) \rmi \gamma_5 \big]
\notag \\
&=
  \frac{\, 2 g^2 \Nc \,}{\, (4\pi)^2 \epsilon \,} \bigg[ M_f^2 + M_g^2 + \big[\, (M_f - M_g)^2 - p^2 \,\big] \bigg]
   \notag \\
&~~~
+
  \frac{\, 2 g^2 \Nc \,}{\, (4\pi)^2 \,}
   \bigg[ M_f^2 + M_g^2 + M_f^2 \ln \frac{\, \Lambda^2 \,}{\, M_f^2 \,} + M_g^2 \ln \frac{\, \Lambda^2 \,}{\, M_g^2 \,} 
   + \big[\, (M_f - M_g)^2 - p^2 \,\big] \calB_{fg} (p^2) 
   \bigg] \,,
\end{align}
with a function including the Feynman parameter,
\beq
 \calB_{fg} (p^2) 
 \equiv  \int_0^1 \! \rmd x ~  \ln \frac{\, \Lambda^2 \,}{\, (1-x) M_f^2 + x M_g^2 - p^2 x(1-x) \,}   \,.
\eeq
This logarithm in the integral can be negative for $p^2 > (M_f + M_g)^2$ and $p^2 < (M_f - M_g)^2$.
(Actually the latter regime does not occur for $0 \le x \le 1$.)
In such case the imaginary part appears as $\ln(- y -\rmi \epsilon) \rightarrow \ln |y| - \rmi \pi$.
The imaginary part describes the decay of mesons into constituent quarks,
but this is supposed to be an artifact of not taking account the quark confinement.
Realistically a meson decays into mesons and the width is typically smaller than the mass, except for some of scalar mesons.
For simplicity, in this work we estimate a meson mass as the location where the real part becomes zero.
Also we discuss the mixing between the flavor singlet and octet using the real part only.

Remember that we chose $\Lambda = M_0$ for the effective potential so we do the same in the following.
For the pseudoscalar channels,
\beq
&&
\hspace{-0.8cm}
\Sigma_\pi^q = 2 \Sigma^q_{ll} \,,~~~~~~ 
\Sigma_K^q = 2 \Sigma^q_{ls} \,,~~~~~~
\notag \\
&&
\hspace{-0.8cm}
\Sigma_{00}^q = \frac{2}{\, \Nf \,} \big( 2 \Sigma_{ll}^{q} + \Sigma_{ss}^q \big) \,,~~~~~~
\Sigma_{88}^q = \frac{2}{\, \Nf \,} \big(  \Sigma_{ll}^{q} + 2\Sigma_{ss}^q \big) \,,~~~~~~
\Sigma_{08}^q = \frac{\, 2\sqrt{2} \,}{\, \Nf \,} \big( \Sigma_{ll}^{q} - \Sigma_{ss}^q \big) \,,
\eeq
Moreover, to obtain the similar expressions for the scalar channel, we can simply replace $(M_f - M_g)^2$ with $(M_f + M_g)^2$.

Let us see some examples. For pions
\begin{align}
\Sigma_\pi^q (p^2)
&=  \frac{\, 4 g^2 \Nc \,}{\, (4\pi)^2 \epsilon \,} \big( 2 M_0^2 - p^2 \,\big)
+
  \frac{\, 4 g^2 \Nc \,}{\, (4\pi)^2 \,}
   \bigg( 2 M_0^2  - p^2 \, \calB_{ll} (m_\pi^2) 
   \bigg) \,.
\end{align}
Combining this expression with the counter term $\Sigma_\pi^{c.t.}$, 
the UV divergence proportional to $M_0^2$ is cancelled by the previously determined $\delta \lambda$.
The divergence coupled to $p^2$ can be canceled by choosing
\beq
\delta Z_\phi = - \frac{\, 4 g^2 \Nc \,}{\, (4\pi)^2 \epsilon \,} \,.
\eeq
Below we write $\Sigma = \Sigma^q + \Sigma^{c.t.}$ for the renormalized self-energy.
The pole condition for pions is $m_\pi^2 =  \big( m_\pi^{\rm tree} \big)^2 + \Sigma_\pi (m_\pi^2) $ with
\beq
\Sigma_\pi (m_\pi^2)
=  \frac{\, 4 g^2 \Nc \,}{\, (4\pi)^2 \,}
   \bigg( 2 M_0^2  - m_\pi^2 \, \calB_{ll} (m_\pi^2) 
   \bigg) \,.
\eeq
For kaons, the above chosen counter terms cancel the divergence in the kaon self-energy,
leaving the condition $m_K^2 =  \big( m_K^{\rm tree} \big)^2 + \Sigma_K (m_K^2) $ with
\begin{align}
\Sigma_K (m_K^2)
= \frac{\, 4 g^2 \Nc \,}{\, (4\pi)^2 \,}
   \bigg( M_0^2 + M_{s0}^2 + M_{s0}^2 \ln \frac{\, M_0^2 \,}{\, M_{s0}^2 \,}
    + \big[ (M_0 - M_{s0} )^2 - m_K^2 \big] \calB_{ls} (m_K^2) 
   \bigg) \,.
\end{align}
We do the same for $00$, $88$, and $08$ channels which are to be diagonalized to yield the $\eta$ and $\eta'$ spectra.
The $\eta$ and $\eta'$ masses are found by searching $p^2$ satisfying 
\beq
\det
\begin{bmatrix}
~ - p^2 + ( m_{00}^{\rm tree} )^2 + \Sigma_{00} (p^2) ~&~ ( m_{08}^{\rm tree} )^2 + \Sigma_{08} (p^2) ~~ \\
~ ( m_{80}^{\rm tree} )^2 + \Sigma_{80} (p^2) ~ & ~ - p^2 + ( m_{88}^{\rm tree} )^2 + \Sigma_{88} (p^2)~~
\end{bmatrix}
= 0\,.
\eeq
The mixing is induced by the flavor breaking associated with $M_l - M_s$.
It is worth mentioning that the mixing angle depends on the energy, not a constant \cite{Hatsuda:1994pi}.

\subsection{Parameter fixing}
\label{sec:parameter_fixing}

We have $(g, m_M^2, \lambda, K, h_l, h_s, M_0, M_{s0})$ for our parameters.
We treat $M_0$ and $M_{s0}$ as if our input parameters,
and set
\beq
M_0 = 0.3\, {\rm GeV}\,,~~~~ M_{s0} = 0.5\, {\rm GeV} \,.
\eeq
Next, we fix $h_l$ and $h_s$ so that the minima of the effective potential in vacuum
are found at $M_l = M_0$ and $M_s = M_{s0}$,
\beq
\frac{\, \partial \Omega \,}{\, \partial M_l \,} \bigg|_{ M_l=M_0,\, M_s = M_{s0} } 
= \frac{\, \partial \Omega \,}{\, \partial M_s \,} \bigg|_{M_l=M_0,\, M_s = M_{s0} } 
= 0 \,.
\eeq
The leftover parameters are $(g, m_M^2, \lambda, K)$.
We have four meson spectra, $(m_\pi, m_K, m_{\eta}, m_{\eta'}) \simeq (0.14, 0.50, 0.55, 0.96 )$ GeV, so the four parameters can be fixed.
Here we use the pseudo-scalar mesons for the parameter fixing instead of using $\sigma$ mesons
since the nature of $\sigma$-mesons are more uncertain than the pseudo-scalar mesons.
The results are summarized in Table.~\ref{tab:coupling} and \ref{tab:mesons} including spectra of scalar mesons.
The $\sigma$ and $\eta'$ masses are sensitive to the value of $K$;
$m_{\sigma}$ is reduced while $m_{\eta'}$ is enhanced by the anomaly.


\begin{table}[H] 
\caption{The parameters in the quark-meson model ($M_0, M_{s0}$ are given in GeV unit).
The parameters are chosen to reproduce the masses of the pseudo scalar nonet.
We use the set B unless otherwise stated.
\label{tab:coupling}}
\newcolumntype{C}{>{\centering\arraybackslash}X}
\begin{tabularx}{\textwidth}{ccccccccc}
\toprule
set  & $M_0$ & $M_{s0}$ & $g$ & $\lambda$ & $ m_M^2\, [{\rm GeV}^2]$  &~ $h_l\, [{\rm GeV}]$ & $h_s\, [{\rm GeV}]$ & $K\, [{\rm GeV}]$  \\
\midrule
A & 0.27 & 0.50 & 3.0 & 38.1 & $-0.269$  & $1.82\cdot 10^{-3}$  & $4.16 \cdot 10^{-2}$  & 1.2   \\
B & 0.30 & 0.50 & 3.3 & 42.4 & $-0.298$  & $1.84\cdot 10^{-3}$  & $4.19 \cdot 10^{-2}$  & 1.6   \\
C & 0.33 & 0.50 & 3.6 & 60.0 & $-0.278$  & $1.85\cdot 10^{-3}$  & $4.32 \cdot 10^{-2}$  & 2.0   \\
\bottomrule
\end{tabularx}
\end{table}

\begin{table}[H] 
\caption{The masses of the pseudo scalar and scalar nonets with the masses indicated in the parenthesis.
(Experimentally the $\sigma$ and $\kappa$ have the width of $\sim 500$ MeV.) 
\label{tab:mesons}}
\newcolumntype{C}{>{\centering\arraybackslash}X}
\begin{tabularx}{1.0\textwidth}{cccccccc}
\toprule
 set  & $m_\pi (138) $ & $m_K (496) $ & $m_\eta (548)$  & $m_{\eta'} (958)$ & $m_\sigma (500^*)$ & $m_{a_0} (980)$ & $m_{\kappa} (700^*)$ \\
\midrule
A  &  0.14 & 0.50 & 0.52  & 0.97  & 0.51 & 0.67  &  0.75 \\
B  &  0.14 & 0.50 & 0.54  & 0.97  & 0.54 & 0.73  &  0.80 \\
C  &  0.14 & 0.50 & 0.55  & 0.95  & 0.64 & 0.81  &  0.88 \\
\bottomrule
\end{tabularx}
\end{table}

\section{Equations of state at zero temperature }
\label{sec:eos}

\begin{figure}[t]
\vspace{-.0cm}
\begin{center}
\includegraphics[width=9. cm]{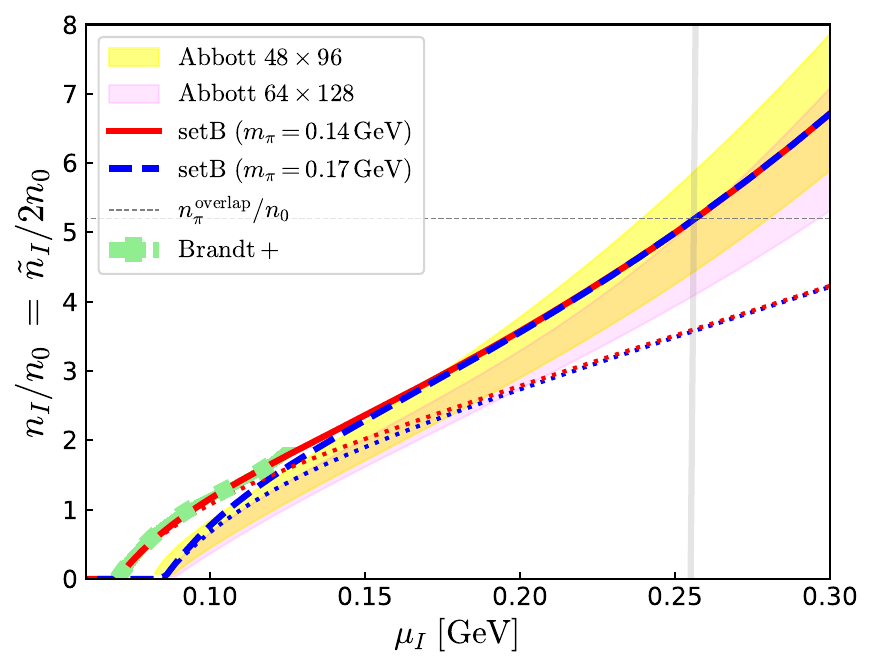}
\end{center}
\vspace{-0.0cm}
\caption{ The isospin density $n_I/n_0 (= \tilde{n}_I/2n_0)$ as a function of $\mu_I$. 
The results of quark-meson model for $m_\pi=0.14$ GeV and $m_\pi = 0.17$ GeV 
are compared with the lattice data by Brandt et al.~\cite{Brandt:2022hwy} and Abbott et al.~\cite{Abbott:2023coj}, respectively.
The overlap density $n_\pi^{\rm overlap} \simeq 5.2n_0$ which is achieved at $\mu_I \simeq 0.256$ GeV is shown as a guide.
The red and blue dotted curves are from the leader order ChPT for $m_\pi=0.14$ and $0.17$ GeV, respectively.
 }
\label{fig:nI-muI}
\end{figure}   

\begin{figure}[t]
\vspace{-.0cm}
\begin{center}
\includegraphics[width=9. cm]{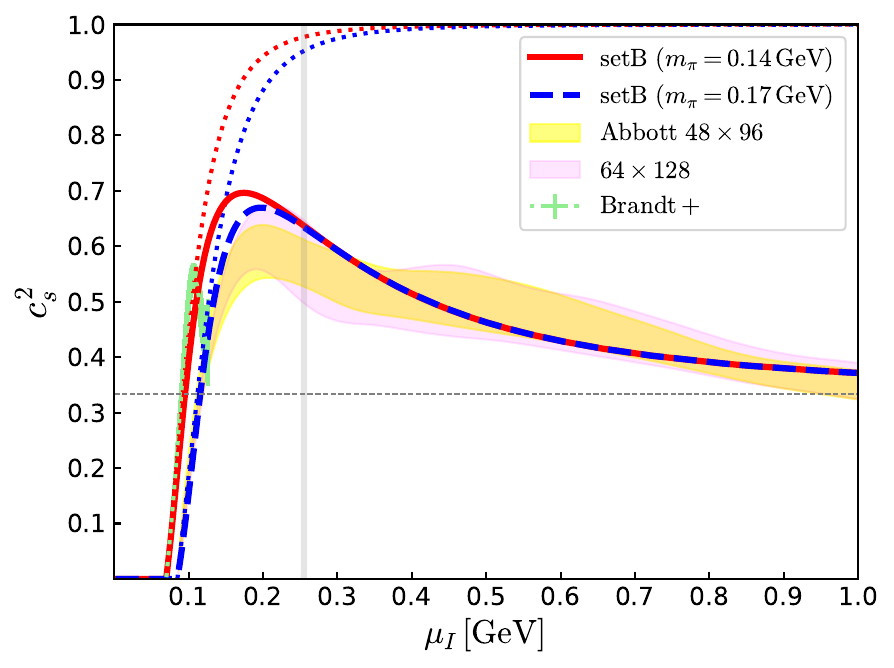}
\end{center}
\vspace{-0.0cm}
\caption{ The sound speed as a function of $\mu_I$. 
The results of quark-meson model for $m_\pi=0.14$ GeV and $m_\pi = 0.17$ GeV 
are compared with the lattice data by Brandt et al.~\cite{Brandt:2022hwy} and Abbott et al.~\cite{Abbott:2023coj}, respectively.
The conformal value $c_s^2=1/3$ and the ChPT results (dotted curves) are also shown as guides.
 }
\label{fig:cs2}
\end{figure}   

\begin{figure}[t]
\vspace{-.0cm}
\begin{center}
\includegraphics[width=9. cm]{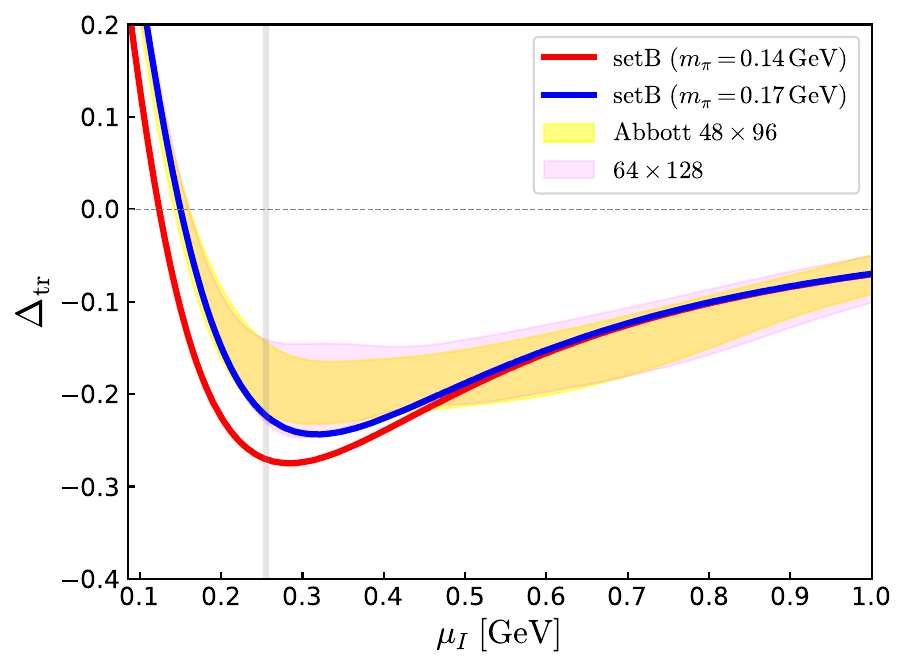}
\end{center}
\vspace{-0.0cm}
\caption{ The trace anomaly $\Delta_{\rm tr}$ for the same setup as Fig.~\ref{fig:cs2}.
The conformal limit is $\Delta_{\rm tr} = 0$.
 }
\label{fig:trace}
\end{figure}   

We examine equations of state.
Substituting the solutions of gap equations into $\Omega$,
the thermodynamic potential is $P = - \Omega$.
The isospin density is
\beq
n_I \equiv \frac{\, \tilde{n}_I \,}{\, 2 \,} =  \frac{\, 1 \,}{\, 2 \,}  \frac{\, \partial P \,}{\, \partial \mu_I \,}  \,.
\eeq
As we mentioned before, we defined $\mu_I (=\mu_u=-\mu_d)$ in unconventional way to simplify the notation.
For physical interpretation we should use $n_I$, rather than $\tilde{n}_I$ directly derivable from the derivative of $P$.
The energy density is $\varepsilon = \mu_I \tilde{n}_I - P$.

At low density the condensate is driven by pions with $I_3=1$;
as far as isospin density is saturated by those pions, one can interpret the isospin density as the pion density, $n_I \simeq n_\pi$.
A good measure to estimate the density where pions overlap is
\beq
n_\pi^{\rm overlap} \equiv \bigg( \frac{\, 4\pi r_\pi^3 \,}{3} \bigg)^{-1} 
\simeq 0.83\, {\rm fm}^{-3} \simeq 5.2 n_0 \,,
\eeq
where we introduced $n_0 = 0.16\, {\rm fm}^{-3}$ as a unit frequently used in nuclear physics.
The estimate utilizes the root-mean-square radius of a pion, $r_\pi \simeq 0.66$ fm, in vacuum \cite{PhysRevD.98.030001,Ananthanarayan:2017efc,Koponen:2015tkr,Wang:2020nbf}.
The size of a pion are comparable to that of a nucleon;
the nucleon charge root-mean-square radius is $\simeq 0.84$ fm,
while its mass radius based on the gravitational form factor \cite{Polyakov:2002yz} 
is $\simeq 0.55$ fm \cite{Kharzeev:2021qkd}
or
the ``valence quark core radius'' is $\simeq 0.5$ fm \cite{Kaiser:2024vbc}.
The relation between $\mu_I$ and $n_I$ is shown in Fig.~\ref{fig:nI-muI}.
The overlap density $n_\pi^{\rm overlap}$ is reached at low chemical potential, $\mu_I \simeq 0.256$ GeV.

A useful quantity characterizing the rapid stiffening is the sound speed.
A matter is called stiff (soft), i.e., $P$ is large (small) at a given $\varepsilon$,
and its variation is reflected in $c_s^2$.
Writing the susceptibility as $\tilde{\chi} = \partial^2 P/\partial \mu_I^2$,
the sound speed is given by 
\beq
c_s^2 
= \frac{\, \rmd P \,}{\, \rmd \varepsilon \,} \bigg|_{s=0}
= \frac{\, \tilde{n}_I \rmd \mu_I \,}{\, \mu_I \rmd \tilde{n}_I \,}
= \frac{\, \tilde{n}_I  \,}{\, \mu_I \tilde{\chi}_I \,} \,.
\eeq
The result of the quark-meson model is shown in Fig.~\ref{fig:cs2}.
The $c_s^2$ grows quickly just after the onset of the pion condensation,
makes a peak before $n_I$ reaching $n_\pi^{\rm overlap}$,
and slowly relaxes to the conformal limit $1/3$.
The result is compared with the lattice results of Brandt et al.~\cite{Brandt:2022hwy} 
which have more focus on low density 
and those of Abbott et al.~\cite{Abbott:2023coj} which study the global structure for wider range of $\mu_I$.
The results are consistent.
Later we delineate the behavior of $c_s^2$ at low and high densities from microphysics viewpoint.

Recently the trace anomaly of EOS attracts attention \cite{Fujimoto:2022ohj,Marczenko:2022jhl,Ma:2019ery,Tajima:2024mvs}. 
It is one of hot topics for forthcoming electron ion collider experiments which will explore the nucleon structure in great detail.
The trace anomaly normalized by $3\varepsilon$ \cite{Fujimoto:2022ohj},
\beq
\Delta_{\rm tr} \equiv \frac{\, \la T^\mu_{\mu}\ra \,}{\, 3 \varepsilon \,} = \frac{1}{\, 3 \,} - \frac{\, P \,}{\, \varepsilon \, }
\eeq
is a useful measure for the deviation from the conformal limit.
The result is shown in Fig.~\ref{fig:trace}.
In the non-relativistic limit, $P \ll \varepsilon$, the $\Delta_{\tr} = 1/3$.
In the conformal limit $\Delta_{\rm tr} = 0$.
When $\Delta_{\rm tr} < 0$, 
there should be strong correlation effects.
Below we discuss the power corrections as candidates to drive $\Delta_{\rm tr}$ to the negative value.

As a guide for the low density behaviors, 
it is useful to quote the results from the chiral perturbation theory (ChPT),
\beq
P = 2 f_\pi^2 \mu_I^2 \bigg( 1 - \frac{\, m_\pi^2 \,}{\, 2\mu_I \,} \bigg)^4 \,,~~~~~ 
\eeq
which leads to
\beq
n_I = \frac{\, \tilde{n}_I \,}{\, 2 \,} 
	= 2f_\pi^2 \mu_I \bigg[ 1 - \bigg(\frac{\, m_\pi \,}{\, 2 \mu_I\,} \bigg)^4 \bigg] \,,~~~~~~~~
c_s^2 = \frac{\, (2 \mu_I)^4 - m_\pi^4 \,}{\, (2 \mu_I)^4 + 3 m_\pi^4 \,} \,.
\eeq
At large density the sound speed asymptotically approaches $c_s^2 = 1$.
Here the scaling $\varepsilon \sim n_I^2$ holds; 
the system is as if dominated by repulsive two-body forces.

\subsection{ High density regime: conformal behaviors }
\label{sec:high_density_conformal}

Let us first consider the high density regime in our model and the effective degrees of freedom.
In the conformal limit with no mass scale other than $\mu_I$,
the dimensional analyses lead to
\beq
\frac{\, \partial (P/\mu_I^4) \,}{\, \partial \mu_I \,} = 0 ~~ \rightarrow ~~ P = \frac{\, \varepsilon \,}{3} \,.
\eeq
This relation is approximately satisfied in the high density limit where the kinetic energy dominates over mass and interaction energies.
(Strictly speaking, there are self-energies that do not necessarily die out at high energy; in QCD the asymptotic freedom guarantees the self-energies become weaker at high energy.)

The conformal limit by itself does not tell which degrees of freedom is relevant.
Indeed, supposing $P \simeq a \mu_I^4$, the conformal limit is obtained for whatever value of $a$;
we have to specify $a$ to tell which degrees of freedom dominates.
As we saw, the tree level pressure including only mesonic degrees of freedom,
and the one-loop pressure (dominated by quark degrees of freedom)
scales as
\beq
P_{\rm tree} \sim \frac{\, 24 \,}{\, \lambda \,} \mu_I^4 \,,~~~~~~~~
P_{\rm 1-loop} \sim \frac{\, \Nf^{\rm eff} \Nc \,}{\, 12 \pi^2 \,} \mu_I^4 \,,
\eeq
where $\mu_I = \mu_u = - \mu_d$ and we should substitute $\Nf^{\rm eff} = 2$.
In order for the former to reproduce the latter, we need $\lambda \simeq 474$ which is unacceptably large.
For $P_{\rm tree}$, the $\Delta_{\rm tree} \sim \mu_I g/\sqrt{\lambda}$ plays an essential role for the conformal scaling 
and its magnitude is determined by details of interactions.
In stark contrast, the conformal scaling of $P_{\rm 1-loop} $ is determined by the number of quark degrees of freedom
and robust to the interactions.

\subsection{ High density regime: perturbative and power corrections }
\label{sec:high_density_pert_power}

Next we consider corrections to the conformal limit.
The most frequently and rigorously discussed is the perturbative corrections at weak coupling regime.
The perturbative corrections modify the coefficient of $\mu_I^4$ term by adding series of $\alpha_s (\bar{\Lambda})$
where $\bar{\Lambda}$ is the renormalization scale to be set to the natural scale for a given problem.
At large density $\bar{\Lambda} \sim \mu_I$ is the natural choice.
The intrinsic scale of QCD, $\lqcd \simeq 200$-$300$ MeV, appearing only through $\alpha_s$;
it shows up in the form of $\sim \ln (\mu_I/\lqcd)$. 
To the one-loop $\beta$-function,
\beq
\frac{\, \partial \alpha_s (\mu_I) \,}{\, \partial \ln \mu_I \,}
= - \frac{\, 11\Nc - 2 \Nf \,}{\, 24\pi \,} \alpha_s^2 \,,~~~~~~\Nf = 3 \,.
\eeq
This reduces $P$, $n_I$, and $c_s^2$ from the conformal value.
At asymptotic density, the $c_s^2$ approaches the conformal limit from below.

In addition, in QCD there are power corrections in powers of $\lqcd/\mu_I$.
This sort of terms cannot be expressed by powers of series of $\alpha_s$,
\beq
\lqcd^{\rm 1-loop} \simeq \mu_I \rme^{- 2\pi/\beta_0 \alpha_s (\mu_I) } \,,~~~~~~~ \beta_0 = 11- 2 \Nf/3 \,.
\eeq
and hence is non-perturbative.
For momentum transfer of $\sim \lqcd$, quarks are supposed to strongly interact.

Actually, whether such soft interactions are important or not depends on the presence of the color screening.
If the screening is strong in the infrared soft interactions are cut off.
In the case of pion condensed phase, the condensate is color-singlet and produces a gap for a quark-hole excitation.
This suppresses the Debye and Meissner screening.
Hence the electric and magnetic color interactions can be as strong as in the vacuum
until the phase space factor $\sim 4\pi \mu_I^2$ overwhelms the suppression factor by the gap.

Adding the power corrections can change the qualitative behavior of $c_s^2$ \cite{Kojo:2014rca,Kojo:2021hqh,Chiba:2023ftg,Geissel:2024nmx,Braun:2022jme,Leonhardt:2019fua}.
In particular the power corrections adding positive contributions to the pressure
favors the trend opposite to those from perturbative corrections,
and hence can be thought of a clear indicator for non-perturbative physics in the high density domain.
For a simple parametrization \cite{Chiba:2023ftg}
\beq
P = a \big( \mu_I^4 + c_2 \Delta^2 \mu_I^2 \big) - B \,, 
\label{eq:P_parametrization}
\eeq
with $a , c_2 >0$ being dimensionless constants and $B$ included for normalization,
we find
\beq
\tilde{n}_I = 2 a \big( 2 \mu_I^3 + c_2 \Delta^2 \mu_I \big) \,,~~~~~
\tilde{\chi}_I = 2 a \big( 6 \mu_I^2 + c_2 \Delta^2 \big) \,,
\eeq
with which
\beq
c_s^2 = \frac{\, 2 \mu_I^2 + c_2 \Delta^2 \,}{\, 6 \mu_I^2 + c_2 \Delta^2 \,} 
= \frac{1}{\, 3 \,} + \frac{2 }{\, 3 \,} \frac{\, c_2 \Delta^2 \,}{\, 6 \mu_I^2 + c_2 \Delta^2 \,} 
\,,
\eeq
varying between 1/3 and 1.
In our quark-meson model $a = \Nc /6\pi^2$, $c_2 \simeq 3$, and $\Delta \simeq 0.25$-0.3 GeV \cite{Chiba:2023ftg}. 
The power corrections of $O(10\%)$ to the pressure occurs when
\beq
3 (\Delta/\mu_I)^2 \sim 0.1 
~\rightarrow~ 
\mu_I \sim 1.6\, {\rm GeV} \times \frac{\, \Delta \,}{\, 0.3\, {\rm GeV} \,} \,,
\eeq
where the corrections increase $c_s^2$ as $0.333 \rightarrow 0.344$ for $\Delta \simeq 0.3$ GeV.
The non-perturbative effects would survive at relatively high density of $\mu_I \sim 1$ GeV.
These analyses, although still crude, indicate that the convergence of $\alpha_s$ expansion at high density, 
which seems to be universal,
may not be sufficient to judge the relevance/irrelevance of the non-perturbative physics.

The lattice results for EOS by Abbott et al.~\cite{Abbott:2023coj} to $\mu_I \sim 1.7$ GeV ($\mu_I$ defined in our work is a half of theirs) 
show qualitative deviation from perturbative QCD predictions.
One of plausible sources is the power corrections from the $\sim \mu_I^2 \Delta^2$ terms where $\Delta$ is non-analytic in the strong coupling constant $g_s$.
This indicates that non-perturbative physics can survive to very high density even where $\alpha_s(\mu_I)$ is reasonably small \cite{Chiba:2023ftg}.
Recently Ref.~\cite{Abbott:2023coj} used the deviation between the perturbative EOS and lattice results to estimate 
the size of the BCS gap in the weak coupling region, $\mu_I \in [0.750, 1.625]$ GeV \cite{Abbott:2024vhj}.
The expression used for the EOS is \cite{Fujimoto:2023mvc}
\beq
\delta P 
= P (\Delta ) - P(\Delta =0)
= \frac{\, \Nc \,}{\, 2\pi^2 \,} \mu_I^2 \Delta^2 \bigg( 1 + \frac{\, g_s \,}{\, 6 \,} \bigg) \,.
\label{eq:gap_from_eos}
\eeq
The lattice results show that $\Delta$ increases toward the low density region.
Meanwhile the gap estimated by the weak coupling method is
\beq
\Delta = \tilde{b} \mu_I  \exp\bigg[- \frac{\, \pi^2 + 4 \,}{16} \bigg] \exp\bigg[- \frac{\, 3 \pi^2\,}{\,2 g_s \,} \bigg] \,,
~~~~~~~~
\tilde{b} = 512 \pi^4 g_s^{-5} \,,
\label{eq:gap_from_weak}
\eeq
which is sensitive to the choice of the running $g_s (\bar{\Lambda})$. 
The estimates with $\bar{\Lambda} = \mu_I$ and $2\mu_I$ covers the range consistent with those estimated from the lattice results and Eq.~\eqref{eq:gap_from_eos}.
On the other hand, 
the $\mu_I$ dependence looks opposite; the gap of Eq.~\eqref{eq:gap_from_weak} decreases toward the low density region.
We conjecture that the physics of soft-gluons and quarks enhances the gap and cures the discrepancy.


The power corrections also play important roles in the trace anomaly.
In the parametrization Eq.~\eqref{eq:P_parametrization}, the energy density is
\beq
\varepsilon = a \big( 3 \mu_I^4 + c_2 \Delta^2 \mu_I^2 \big) + B \,, 
\eeq
so that
\beq
\la T_\mu^\mu \ra 
= \varepsilon - 3P
= - 2 a c_2 \Delta^2 \mu_I^2 + 4 B \,.
\eeq
Remarkably the power corrections yield negative contributions.
Meanwhile the normalization (bag) constant measures the energy difference between the conformal and non-perturbative vacuum 
and should be $\sim \lqcd^4$ and positive.
Perturbative corrections not written here also give positive contributions \cite{Chiba:2023ftg}.
Hence the negative trace anomaly can be regarded as a good indicator for the importance of the non-perturbative physics in dense matter.

\begin{figure}[t]
\vspace{-.0cm}
\begin{center}
\includegraphics[width=9. cm]{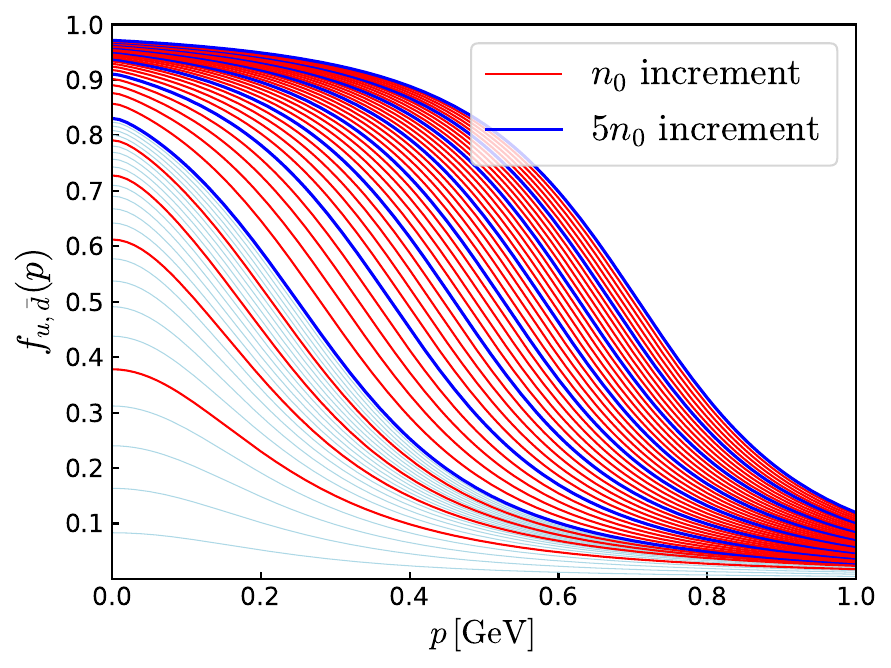}
\end{center}
\vspace{-0.0cm}
\caption{ The occupation probability $f_{u,\bar{d}} (p)$ as a function of $p$ for the parameter set B in Table.~\ref{tab:coupling}.
Red curves represent curves for $n_0$ increments, $n_0, 2n_0, \cdots$
while blue curves for $5n_0$ increments.
To $5n_0$ we also show thin light blue curves for $0.2n_0$ increments.
The figure covers to $n_I = 20n_0$.
 }
\label{fig:fQ_pvary}
\end{figure}   

\begin{figure}[t]
\vspace{-.0cm}
\begin{center}
\includegraphics[width=9. cm]{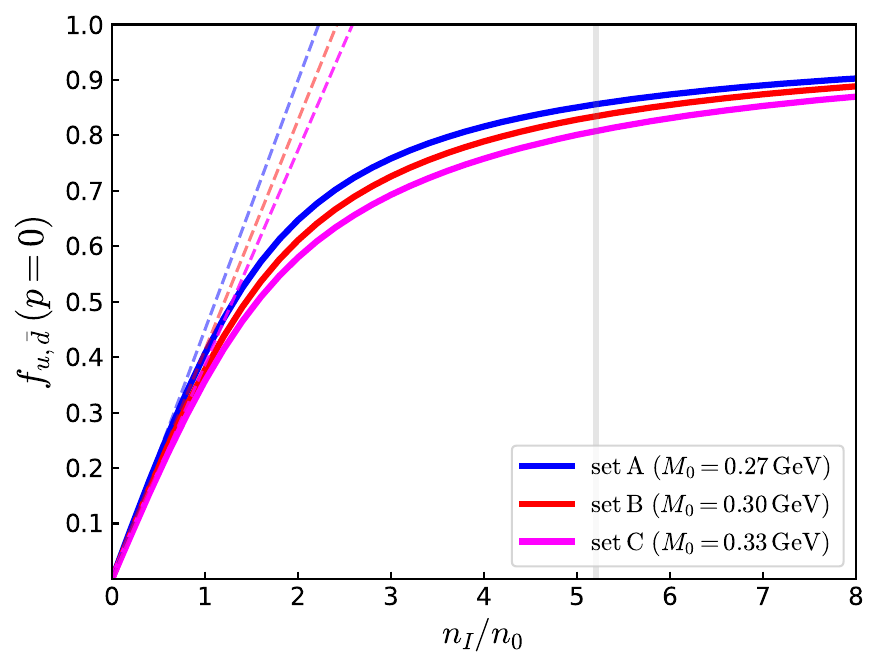}
\end{center}
\vspace{-0.0cm}
\caption{ The occupation probability at zero momentum, $f_{u,\bar{d}} (p=0)$ at various $n_I/n_0$. 
With the parameter sets A-C, we vary $M_0 (=M_{u,d}^{\rm vac})$ to change the ``binding energies'' for quarks to form pions;
a larger $M_0$ has the stronger binding, $\sim 2M_0 - m_\pi$, and leads to a more spatially compact state. 
The dashed curves are linear extrapolation of the low density behavior, $f_{u,\bar{d}}^{\rm id} (n_I) \equiv n_I \partial f_{u,\bar{d}}(p=0)/\partial n_I \big|_{n_I=0}$,
which characterizes the $f_{u,\bar{d}}$ for non-interacting pions.
We also show $n_\pi^{\rm overlap}$ as a guide.
}
\label{fig:fQ_p0}
\end{figure}   

\begin{figure}[t]
\vspace{-.0cm}
\begin{center}
\includegraphics[width=9. cm]{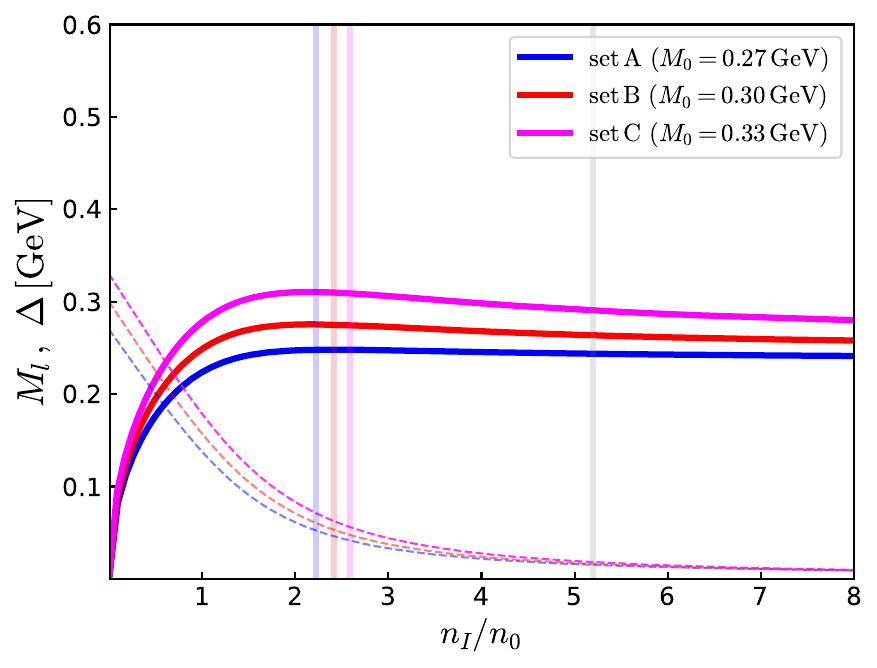}
\end{center}
\vspace{-0.0cm}
\caption{ Behaviors of $M_l$ (dashed lines) and $\Delta$ (solid lines) for the same parameter set as Fig.~\ref{fig:fQ_p0}.
The vertical lines with various colors represent the $n_I^{\rm id}$ defined in Eq.~\eqref{eq:n_id}
for the set A, B, and C.
}
\label{fig:gap_M0vary}
\end{figure}   

\begin{figure}[t]
\vspace{-.0cm}
\begin{center}
\includegraphics[width=9. cm]{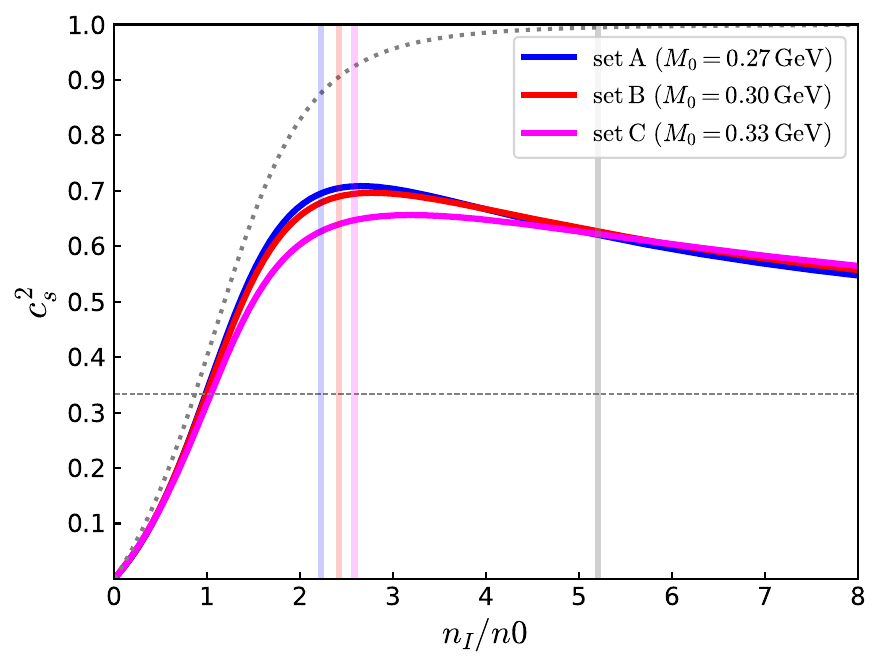}
\end{center}
\vspace{-0.0cm}
\caption{ The sound speed as a function of $n_I/n_0$ for the same parameter set as Fig.~\ref{fig:fQ_p0}.
When pions are more spatially compact ($2M_0 - m_\pi$ is larger),
the repulsion among pions or quark saturation effects set in at larger densities.
}
\label{fig:gap_M0vary}
\end{figure}   

\subsection{ Quark saturation }
\label{sec:quark_saturation}

Finally we try to characterize EOS through quarks inside of pions.
At sufficiently low density quarks should show up only as constituents of pions.
Pion condensation generates BCS gaps for quarks with which quark contributions
can become finite even for $\mu_I (=\mu_u=-\mu_d) < M_l$, see the structure of Eq.~\eqref{eq:bcs_gap}.
In the pion condensed phase the occupation probability of $u,\bar{d}$ quark states can be written as
\beq
f_Q(p;n_I) \equiv f_{u,\bar{d}} (p;n_I) = \frac{1}{\, 2 \,} \bigg( 1 - \frac{\, E_l - \mu_I \,}{\, \sqrt{ (E_l -\mu_I)^2 + \Delta^2 } \,} \bigg) \,,
\eeq
which becomes $\Theta (\mu_I - E_l)$ for $\Delta \rightarrow 0$.
The behavior of $f_Q (p)$ as a function of $p$ for various $n_I$ is shown in Fig.~\ref{fig:fQ_pvary} for the parameter set B.

As quarks are bound to a compact object, the quark momentum distribution is broad in momentum.
In particular the occupation probability at $p=0$ is substantially smaller than 1.
As the density increases, the magnitude of $f_Q$ becomes larger.
If we neglect the interactions and the structural changes of pions,
we would find the scaling
\beq
f_Q^{\rm id} (p;n_I) \equiv n_I \frac{\, \partial f_Q (p;n_I) \,}{\, \partial n_I \,}\bigg|_{n_I=0} \,,
\eeq
where pions condense into zero momentum state;
each pion gives the same quark contribution and hence simply scale as $\sim n_I$.
Obviously extrapolating this expression would violate the Pauli exclusion principle for quarks, $f_Q(p) \le 1$ for any $p$.
We define the ``(pseudo-)quark-saturation'' density $n^{\rm id}_I$ as
\beq
1 = f_Q^{\rm id} (p=0; n_I^{\rm id} ) \,, 
\label{eq:n_id}
\eeq
and use $n^{\rm id}_I$ as the characteristic measure where either pion interactions or quark saturation constraints become important.

Figure~\ref{fig:fQ_p0} shows $f_Q(p=0;n_I)$ as a measure of the quark saturation.
We also show the $f_Q^{\rm id}$ with dashed lines to examine $n_I^{\rm id}$.
To examine how $n_I^{\rm id}$ depends on the compactness, we vary $M_0$ to change the binding energy of a pion, $2M_0 - m_\pi$,
while maintaining spectra of pseudo-scalar nonets within reasonable range.
Concretely we compare the set A-C in Table.~\ref{tab:coupling}.
With a larger $M_0$ the $n_I^{\rm id}$ is larger as the pion is spatially more compact and has broader quark momentum distribution.
For $M_0=(0.27,0.30, 0.33)$ GeV, we found
$n_I^{\rm id} \simeq (2.2, 2.4, 2.6)n_0$, respectively.
These densities are substantially smaller than $n_\pi^{\rm overlap} \simeq 5.2n_0$.

In principle, by specifying $f_Q $ and gaps $(M_l\,, \Delta\,, M_s)$ at a given $\mu_I$,
one can reconstruct the corresponding EOS as
\beq
P (\mu_I) = \int_0^{\mu_I} \rmd \mu' n_I (\mu') = \int_0^{\mu_I} \rmd \mu' \int_{p}  \sum_f \big[ f_{Q}(p;\mu') - f_{\bar{Q}}(p;\mu') \big] \,,
\eeq
hence the evolution of $f_Q$ (and $f_{\bar{Q}}$ for anti-particles) contains sufficient information to study the EOS.

Shown in Fig.~\ref{fig:gap_M0vary} are the behaviors of $M_l$ and $\Delta$ at several densities and $M_0$.
For a greater $M_0$ the gap $\Delta$ is naturally larger.
As guides we display $n_I^{\rm id}$ using several vertical lines.
The general tendency is that, beyond $n_I^{\rm id}$, the gap becomes insensitive to $\mu_I$ 
reflecting that the gap equations are dominated by soft gluon and soft quark contributions.
The parameter set A-C leads to the gap of $\sim 0.23-0.28$ GeV around $\mu_I \sim 1$ GeV.
Hard gluon contributions, which are omitted in this work, should further enhance the size of the gap
and introduce the stronger $\mu_I$ dependence as it is sensitive to the phase space around the Fermi surface.

Finally we examine the sound speed for varying $M_0$.
For smaller $M_0$ rising of $c_s^2$ occurs at lower density.
This is natural as pions are less compact objects.
Increasing $M_0$ delays rising of $c_s^2$.
The structure of a hadron, its valence core size and quantum fluctuations around it \cite{Fukushima:2020cmk},
has the direct relevance to stiffening of matter.
The relation between the baryon structure and nuclear matter has been discussed in, e.g., Refs.~\cite{Saito:2005rv,Geesaman:1995yd,Koch:2024zag,McLerran:2024rvk}.

\section{Summary}
\label{sec:summary}

In this paper we study the EOS of isospin QCD and its relationship to the microphysics.
We used a quark-meson model that interpolates the hadronic and quark sector at microscopic level.
The model is renormalizable so that it can be used to cover from low to high densities, at least formally.
Although the model does not cover the aspects of QCD caused by hard gluons, 
it captures some aspects of the physics caused by soft gluons in a parametrized manner.

One of important issues in dense QCD is how non-perturbative effects relevant in hadron physics die out.
Recent lattice results for EOS \cite{Abbott:2023coj} to $\mu_I \sim 1.7$ GeV ($\mu_I$ defined in our work is a half of theirs) 
indicate the importance of the power corrections from the $\sim \mu_I^2 \Delta^2$ terms with $\Delta$ being non-analytic in the QCD coupling constant $g_s$,
even at high density where  $\alpha_s(\mu_I)$ is reasonably small \cite{Chiba:2023ftg}.
Concerning the size of the gap, there are several questions to be answered.
The first question is at which density the evaluation of the gap is closed within the weak coupling regime;
in the weak coupling estimates we assume that hard momentum transfer processes dominate the gap equation because of the large phase space.
This mechanism is sensitive to the density and at low density soft gluons should become important.
How the transition between these two regimes occur is directly related to the reliability of various estimates.
The second question is to what extent the extrapolation from isospin QCD to QCD at finite baryon density can be valid.
The mechanism for the emergence of the BCS gap is the diquark condensates in the color-superconductivity.
The typical estimate for the gap, mostly based on the weak coupling picture, is $\Delta \lesssim 100$ MeV;
the coefficient of $1/g_s$ in the weak coupling expression of gap is larger than in the isospin QCD so that
the gap is smaller.
Analyses predicting a greater gap is mostly based on effective models constrained by hadron physics.
It is important to fill the gap between two regimes.

The quarks in pions seem important before reaching the overlap density for pions, $n_\pi^{\rm overlap} \simeq 5.2n_0$, 
which is inferred from the size of a pion in vacuum.
Even if we neglect the structural change of pions such as swelling or dissociation,
we cannot go much beyond $\sim 0.5 n_\pi^{\rm overlap} \simeq 2.6n_0$ neglecting constraints from the quark Pauli blocking.
The sound speed peak is also found around $\sim 0.5n_\pi^{\rm overlap}$.
If the quark exchange interactions (or meson exchange) among pions effectively increase the size of pions,
quark states at low momenta get saturated more quickly, inducing the Pauli blocking effects even earlier.
The quark Pauli blocking near hadronic matter should also give insights on many-body forces among hadrons.
In the context of neutron star EOS, it is typical to utilize two- and three-body repulsion
to satisfy the two-solar mass constraints.
But in such descriptions there always remains a question of how to handle the convergence of many-body forces.
We need an organizing principle. 
We guess that the quark saturation effects do the job.

Analyses in this paper left several important problems.
In methodology we should improve one-loop results.
Another important topic is the meson spectra at finite density, including quark loops.
We compute the meson spectra in vacuum for the parameter fixing but have not performed analyses at finite density.
The latter is necessary to answer to interesting questions such as how mesons dissociate and change the structure.
How mesons or quarks in medium contribute to the entropy is also important to understand the color confinement at finite density,
a question originally posed in the quarkyonic matter hypothesis \cite{McLerran:2007qj}.
The analyses toward this issue is in progress.

\funding{
T.K. was supported 
by JSPS KAKENHI Grant No. 23K03377 and No. 18H05407,
and by the Graduate Program on Physics for the Universe (GPPU) at Tohoku university;
D.S. by JSPS KAKENHI Grant No. 23H05439.
}

\acknowledgments{
We thank Drs. Brandt and Endrodi for kindly providing us with their lattice data in Ref.~\cite{Brandt:2022hwy},
and Dr. Abbott and his collaborators for their kindness of sending the lattice data in Ref.~\cite{Abbott:2023coj}.
T.K. thanks Drs. Yuki Fujimoto and Larry McLerran for discussions on the quark saturation effects.
The authors thank the Yukawa Institute for Theoretical Physics at Kyoto University and RIKEN iTHEMS. 
Discussions during the workshop (YITP-T-23-05) on “Condensed Matter Physics of QCD 2024” were useful to complete this work.
}

\conflictsofinterest{The authors declare no conflicts of interest.} 



\abbreviations{Abbreviations}{
The following abbreviations are used in this manuscript:\\

\noindent 
\begin{tabular}{@{}ll}
QC$_2$D & two-color quantum chromodynamics \\
QCD$_I$ & isospin QCD \\
EOS & equations of state \\
$\lqcd \simeq 200$-300 MeV & non-perturbative scale in QCD \\
$n_0 \simeq 0.16\, {\rm fm}^{-3}$ & nuclear saturation density \\
ChPT & Chiral perturbation theory
\end{tabular}
}

%
%
%

\begin{adjustwidth}{-\extralength}{0cm}

\reftitle{References}


\bibliography{ref}

\PublishersNote{}
\end{adjustwidth}
\end{document}